**Global analysis reveals persistent shortfalls and regional differences in availability of foods needed for health.**

Leah Costlow[1*], Anna Herforth[2], Timothy B. Sulser[3], Nicola Cenacchi[3], William A. Masters[1,4]

**Affiliations:**

1. Friedman School of Nutrition Science and Policy, Tufts University

2. Food Prices for Nutrition Project, Boston

3. Foresight and Policy Modeling, International Food Policy Research Institute (IFPRI), Washington D.C.

4. Department of Economics, Tufts University

**Contact author:**

* Leah Costlow, Tufts University, 150 Harrison Avenue, Boston MA 02111 USA

email leah.costlow@tufts.edu

**Acknowledgements:**

This work was funded by UKAid and the Bill & Melinda Gates Foundation through the Food Prices for Nutrition project (INV-016158). We thank our program officer, Shelly Sundberg, as well as many project collaborators for making this work possible. Contributions from IFPRI were also supported by the CGIAR Initiative on Foresight.

**Global analysis reveals persistent shortfalls and regional differences in availability of foods needed for health.**

**Abstract**

Most people around the world still lack access to sufficient quantities of all food groups needed for an active and healthy life. This study traces historical and projected changes in global food systems toward alignment with the new Healthy Diet Basket (HDB) used by UN agencies and the World Bank to monitor the cost and affordability of healthy diets worldwide. Using the HDB as a standard to measure adequacy of national, regional and global supply-demand balances, we find substantial but inconsistent progress toward closer alignment with dietary guidelines, with large global shortfalls in fruits, vegetables, and legumes, nuts, and seeds, and large disparities among regions in use of animal source foods. Projections show that additional investments aimed at reducing chronic hunger would modestly accelerate improvements in adequacy where shortfalls are greatest, revealing the need for complementary investments to increase access to under-consumed food groups especially in low-income countries.

**1. Background and motivation**

International efforts to reduce the burdens of malnutrition and diet-related disease depend on increased supply of the food groups needed for healthy diets. New initiatives are moving beyond historical investments designed to meet dietary energy needs (Pingali, 2012; Pingali, 2015), towards elimination of micronutrient deficiencies and inadequate diets among the most vulnerable (FAO, IFAD, UNICEF, WFP and WHO, 2024; Vos et al., 2020) and prevention of diabetes, hypertension, and other diseases (Abarca-Gómez et al., 2017; Popkin et al., 2012). Poor diets and malnutrition are the world's leading modifiable risk factors for death and disability, with suboptimal diets resulting in 11 million premature deaths globally in 2017 (Afshin et al., 2019).

Governments often use food-based dietary guidelines (FBDGs), produced by national health authorities to summarize the evolving scientific consensus, to define healthy diets and help inform consumers on which food groups to eat and in what proportions to maintain energy balance, meet nutrient requirements, and protect against noncommunicable diseases (FAO, 2023). There is substantial agreement among FBDGs, recommending consumption of fruits, vegetables, starchy staples, animal-source foods, oils and fats, and legumes, nuts and seeds in appropriate proportions (Herforth et al., 2019), and there is clear evidence that dietary patterns around the world fall short of these recommendations (Global Diet Quality Project, 2022; Miller et al., 2022). Previous studies have addressed how global food supplies reach or fall short of requirements for essential nutrients (Lividini & Masters, 2022; Wang et al., 2023), specified food groups (fruits and vegetables, sugar, red and processed meat, fish and seafood) (Clifford Astbury et al., 2021; Mason-D'Croz et al., 2019), or other reference diets including the EAT-Lancet dietary patterns or the Harvard Healthy Eating Plate (Tuninetti et al., 2022; Krishna Bahadur et al., 2018). Each of these studies finds shortfalls in fruits and vegetables, and where examined, shortfalls of other food groups (e.g. legumes and nuts according to Tuninetti et al., 2022, and protein and milk, according to Krishna Bahadur et al. 2018), and excesses of sugar, starchy staples, meat, and dietary energy. However, there is not yet a comprehensive picture of how quantities produced and consumed around the world align with total diets composed of the food groups specified in national dietary guidelines.

Since 2022 the new global Healthy Diet Basket (HDB) has been used by UN agencies and the World Bank for global monitoring of the cost and affordability of a healthy diet (FAOSTAT 2024). The HDB specifies benchmark quantities across six food groups (starchy staples; fruits; vegetables; animal-source foods; legumes, nuts, and seeds; and oils and fats), expressed in terms of dietary energy per day, and representing common requirements among national FBDGs (Herforth et al. 2022, 2024). It is used for monitoring CoHD by the UN agencies because it is the first reference diet that has been explicitly derived from FBDGs that countries around the world have published, reflecting broad international consensus in healthy diet patterns. As in all FBDGs, the food groups are constant as generally universal components of a healthy diet, but there is wide variability in the locally-available food items that can satisfy them (for example, groundnut sauce, dal, and hummus all fit in the legumes, nuts, and seeds group). Therefore within each food group, any combination of items may be consumed according to availability and preferences, allowing for local and regional variation in diets. The HDB composition is broadly similar to all national FBDGs and other reference diets such as the Harvard Healthy Eating Plate (analyzed in Krishna Bahadur et al., 2018), but we use the HDB because it is used by the UN agencies for global monitoring of access to healthy diets since 2022. Diet baskets meeting HDB quantity benchmarks have adequate macronutrient balance and a mean adequacy ratio of 95% for micronutrients and protein for the reference person of a healthy adult woman (Herforth et al., 2024). National guidelines and other recommendations often specify additional targets, such as those for intakes of whole grains, salt, and sugar (Herforth et al., 2019); of these, food supply data are amenable to comparison against international guidelines on sugar consumption (WHO 2015).

      This analysis compares historical and projected data for quantities by from each HDB food group, relative to the HDB reference quantities. For historical data on quantities, we look to FAO food balance sheets since 1961. For future projections, we build on the historical evidence from food balance sheets by incorporating economic modeling projections from the International Model for Policy Analysis of Agricultural Commodities and Trade (IMPACT). IMPACT is a partial equilibrium modeling framework developed by IFPRI to simulate changes in agricultural markets under a variety of scenarios for major food system drivers, such as climate change, technology, and socioeconomic changes. Increased agricultural investments are the primary policy change of interest in the context of improving food supplies, especially in low- and middle-income countries. We use results from an IMPACT modeling exercise designed to estimate the cost of investments required to counteract the rise in chronic hunger that would otherwise occur under climate change, as measured by the prevalence of undernourishment (Sulser et al., 2021).

      This analysis shows the extent to which adequate quantities of food exist in the food supply for all people to meet dietary needs in each country and year, under optimistic conditions of equitable distribution and access within countries across diverse populations and subnational regions. This approach provides the first global analysis of food availability relative to a single benchmark for healthy diets that reflects the international consensus across dietary guidelines.

## 2. Data and Methods

*Food Balance Sheets*

We use food supply data from the FAO Food Balance Sheets, beginning with data from 153 countries in 1961 and ending with data from 188 countries in 2022 (FAO, 2024b, 2024c). Food supply is calculated as the daily per capita energy (using population data compiled by FAO) available in each country and each year, beginning with domestic production and accounting for net imports, pre-retail loss, changes in stocks, and non-food uses such as feed, seed, and industrial applications.

We match each FAO commodity to its corresponding HDB food group and combine some FAO commodities into aggregate groups when the quantity available is small. For example, yams are added to the pre-existing group "other roots", while demersal fish, pelagic fish, and other marine fish are combined into the new group "other marine fish". See Supplementary Table S1 for a complete list of these groupings. We obtain regional aggregates by adding national availability by commodity and HDB food group for each subregion in the United Nations geoscheme and combine subregions to reach 7 total world regions. RStudio (version 2024.04.1+748) was used to combine the old and new FBS databases and to perform all data analysis and visualizations. R code is available to replicate and update the results presented here with future annual updates to the FBS database. Country inclusion as well as starting and ending years for each country are described in Supplementary Table S2, while regional data are shown in Table S3.

*Healthy Diet Basket and Healthy Diet Basket Index (HDBI)*

The Healthy Diet Basket (HDB) specifies reference intakes for six food groups in recommended amounts derived from numerous national FBDGs, calibrated to meet the energy needs of a reference adult woman (Fig. 1). Free sugars are not included in the HDB, so we assess sugar supply using the WHO guideline on limiting free sugar intake to less than 10% of total dietary energy (World Health Organization, 2015).

**Table 1. Healthy Diet Basket (HDB) amounts per capita**

| Food Group | Target intake (kcal/ day) |
|---|---|
| Starchy staples | 1160 |
| Fruits | 160 |
| Vegetables | 110 |
| Animal-source foods | 300 |
| Legumes, nuts, and seeds | 300 |
| Oils and fats | 300 |
| Total | 2,330 |

Source: Herforth et al., 2022

To construct a Healthy Diet Basket Index (HDBI), each country or region's average shortfall below HDB targets summarizes the extent to which national, regional, or global food systems provide sufficient foods to meet recommended quantities in national dietary guidelines. Over the six recommended food groups $i$, we report the average degree to which available quantities $q_i$ fall below the HDB target $Q_i$, expressed as the absolute value of proportional shortfalls from 0 to -1 for food groups where $q_i$ is below $Q_i$. Our index of adequacy is 1 minus the average shortfall over all six food groups:

$$HDBI = 1 - \frac{1}{6}\sum_{i=1}^{6} \left|\frac{q_i - Q_i}{Q_i}\right| \quad \forall\ q_i < Q_i \tag{1}$$

This metric is analogous to the mean adequacy ratio used for essential nutrients, as shortfalls below targets for each food group are equally weighted and not offset by excesses of any food group. The highest attainable adequacy score of 1 represents a benchmark situation with zero shortfalls, while an adequacy score of 0 would be a hypothetical extreme where all energy was obtained from sugar. Possible scores indicating sufficient energy but limited food supply would include 0.167 for a scenario where all energy requirements are met from starchy staples alone, or 0.333 if all energy is from combinations of starchy staples and pulses such as chapati and dal, maize and peanut stew, or rice and beans, for example. A score of 0.833 could arise if four different food groups are used at or above target levels and the two remaining groups are consumed at 50 percent of their HDB target.

The shortfall approach is calculated under an assumption of perfect distribution. This assumption is never true, so in reality a higher amount than the per capita target would be needed for all people to access healthy diets, given that some people will overconsume. However, there is no clear method for determining the buffer amount of each food group that would need to be available in the food supply to account for imperfect distribution. Therefore we use the per capita daily recommended amounts as an absolute lowest bound on adequate supply, with the interpretation that if supplies are below adequate amounts in a perfect distribution scenario, then they are certainly below adequate amounts for all to access healthy diets under conditions of imperfect distribution. The HDBI shortfalls therefore are the most conservative description of shortfalls, and actual shortfalls would be greater by an undefined amount due to imperfect distribution.

*Projected HDBI scores with the IMPACT model*

Full IMPACT model specifications are discussed at length in related publications (Mason-D'Croz et al., 2019; Robinson et al., 2015; Rosegrant et al., 2017; Sulser et al., 2021). In this study we include two scenarios which capture possible future investments in agricultural R&D. The "reference" (or business as usual) scenario projects forward historical investment trends made by national agricultural research systems and the research centers unified within the Consultive Group on International Agricultural Research (CGIAR), resulting in an estimated 2.5 to 3 percent annual growth in investments. The "increased investments" scenario models increased CGIAR investments to improve agricultural water use and increase supply and reduce prices of specific commodities in some food groups (not including vegetables or fruits other than bananas). These include investments in (1) international research and development (R&D) for genetic improvement and accompanying management changes in these commodities, in proportion to past R&D investments, (2) infrastructure and institutional change for reduced marketing margins, (3) expanded and more efficient irrigation, and (4) technologies leading to improved soil quality through better water-holding capacity (for individual scenarios see Supplementary Figure S1). Agricultural R&D investments are calibrated to close yield gaps for cereals, roots, tubers, plantains and bananas, animal-sourced foods, pulses, and oilseeds. Subsequent changes in other food groups are modeled from increased incomes and other compounding effects linked to increased productivity in the targeted crops. The investments are focused on target countries for the CGIAR and therefore exclude high-income regions.

Both scenarios are modeled under pessimistic conditions for both climate impacts and socioeconomic outcomes. The RCP8.5 climate change scenario assumes limited reductions in greenhouse gas emissions, and the SSP3 or "Regional Rivalry" shared socioeconomic pathway assumes relatively high population growth and relatively low income growth. Under these near-worst-case conditions, the IMPACT model estimates the magnitude of investments required to prevent new increases in chronic hunger in CGIAR focus countries. The model holds current trends constant in most wealthy countries and estimates the results of increased agricultural investments in most low- and middle-income countries. We calculate the percentage change in each food group's availability under each investment scenario and estimate HDBI scores by applying projected changes to the observed FBS availability of each food group in each region in the IMPACT reference year of 2010.

A total of 158 countries and country aggregates are included in the IMPACT model. Other than for a few aggregate regions (e.g. for small island regions), there is a close correspondence between IMPACT and FBS country inclusion. For Figure 4, we use 2010 population data to calculate food group totals for IMPACT country aggregates when all countries within an IMPACT country group are included in FBS data. This analytical step is possible for the IMPACT country aggregates Baltic States, Other Balkans, Other Southeast Asia, and Rest of Arab Peninsula, but is not possible for country aggregates including numerous small island territories (see Supplementary Table 5).

## 3. Results

*Past trends*

Food systems provide varying levels of total dietary energy, as shown in Table 2. These data represent the sum of each population's food use, from estimated production plus imports and minus exports, non-food uses and losses prior to household consumption. Data are transformations of FAO Food Balance Sheet estimates for 1961 to 2022 from national government data (FAO, 2024b, 2024c), adjusted for region population and expressed in terms of daily energy available per capita in each year. Totals shown in the table exceed dietary intake due to kitchen and plate waste or other uses, and dietary intake itself varies with height, weight and physical activity.

**Table 2. Daily energy available per capita globally and by region, 1960s and 2020s**

| Region | 1960s | 2020s |
|---|---|---|
| East Asia & Pacific | 1,822 | 3,096 |
| Europe & Central Asia | 2,963 | 3,224 |
| Latin America & Caribbean | 2,278 | 2,986 |
| North America | 2,792 | 3,691 |
| South Asia | 2,004 | 2,532 |
| Sub-Saharan Africa | 1,982 | 2,385 |
| Western Asia & North Africa | 2,242 | 3,066 |
| World | 2,225 | 2,881 |

Note: Data shown are the mean regional and worldwide totals from 1961 through 1969 (1960s) and from 2020 through 2022 (2020s), adding up total supply including the six HDB food groups plus sugars. Remaining commodities including spices, infant foods, beverages, and stimulants are excluded from these totals.

The composition of aggregate global food supplies in terms of the six HDB food groups is shown in Figure 1.

**Figure 1. Global food supply relative to Healthy Diet Basket targets, 1961-2022**

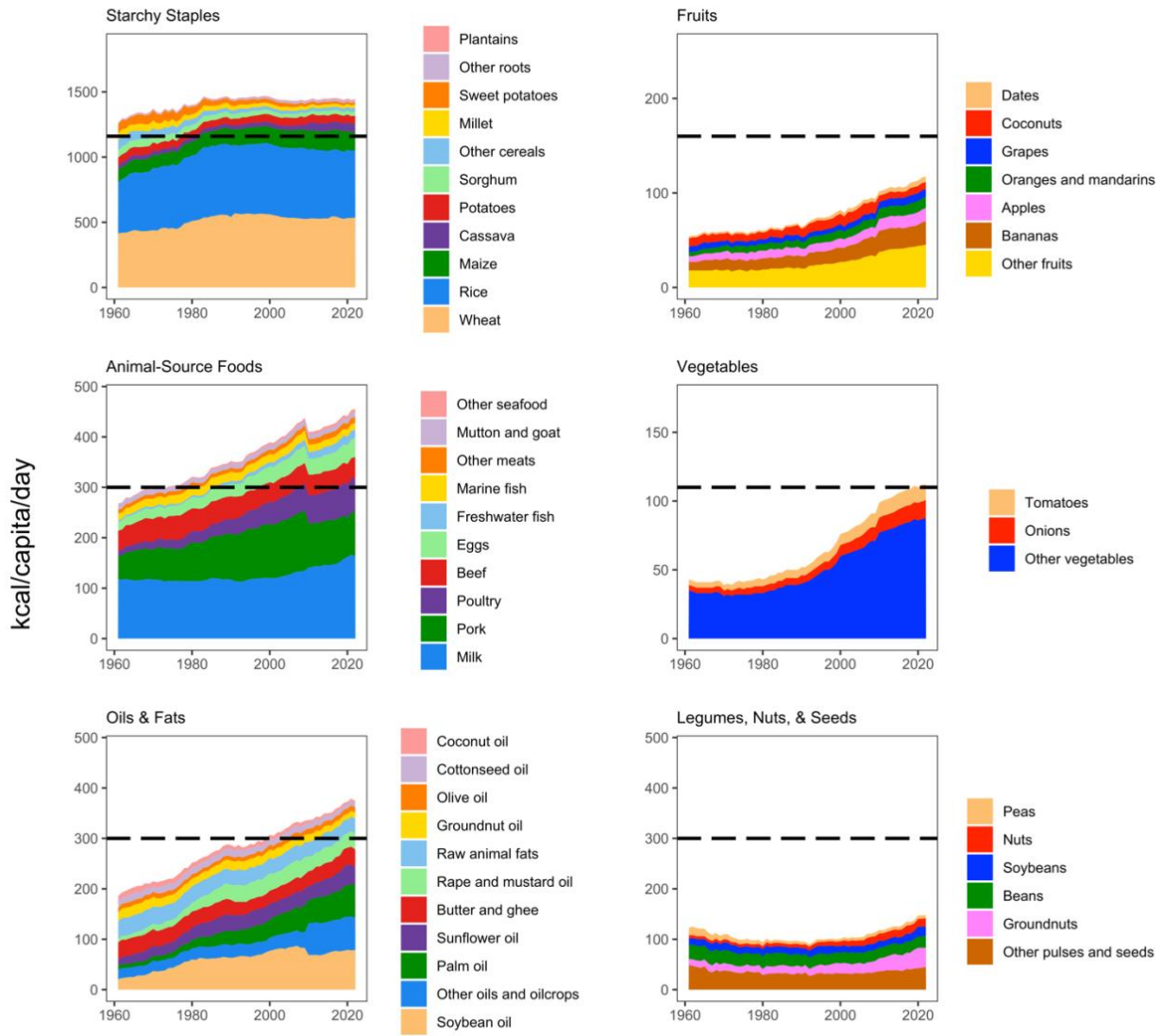

Note: Data shown are worldwide totals for food supply (kcal/capita/day), computed by the authors from national data in FAOSTAT. Plots are scaled to reference intake values. Black dashed lines indicate Healthy Diet Basket targets used to measure access to a healthy diet as recommended in national dietary guidelines.

The per capita global availability of all six food groups has increased in different ways. For starchy staples, global supplies rose above the target for a balanced diet in the 1960s and 1970s but have not increased since the mid-1980s. The HDB amount of animal source foods was reached in the 1970s and supply has continued to rise, exceeding the target amount. The supply of oils and fats has continued to rise beyond the HDB target, which was reached in the early 2000s. Fruit and vegetable supplies saw little or no increase in the 1960s and 1970s but have grown sharply since then, especially for vegetables, while supplies of legumes, nuts, and seeds actually declined through the 1960s and 1970s before stabilizing in the 1980s and 1990s and rising slowly since the 2000s. Additionally, the composition of each food group has changed over time in important ways. For starchy staples, wheat and rice remain dominant, while animal source foods are increasingly provided by milk, poultry, and pork. The most dramatic shift is the increasing role of palm and soybean oil, as well as bananas among fruits, onions, tomatoes, and other vegetables. Among legumes, nuts and seeds, groundnuts have increased but the diverse groups of "beans" and "other pulses and seeds" have declined in per capita availability since 1961.

Some trends in food availability by food group have been similar among major world regions, with important differences as shown in Figure 2.

**Figure 2. Regional food supplies relative to Healthy Diet Basket targets, 1961-2022**

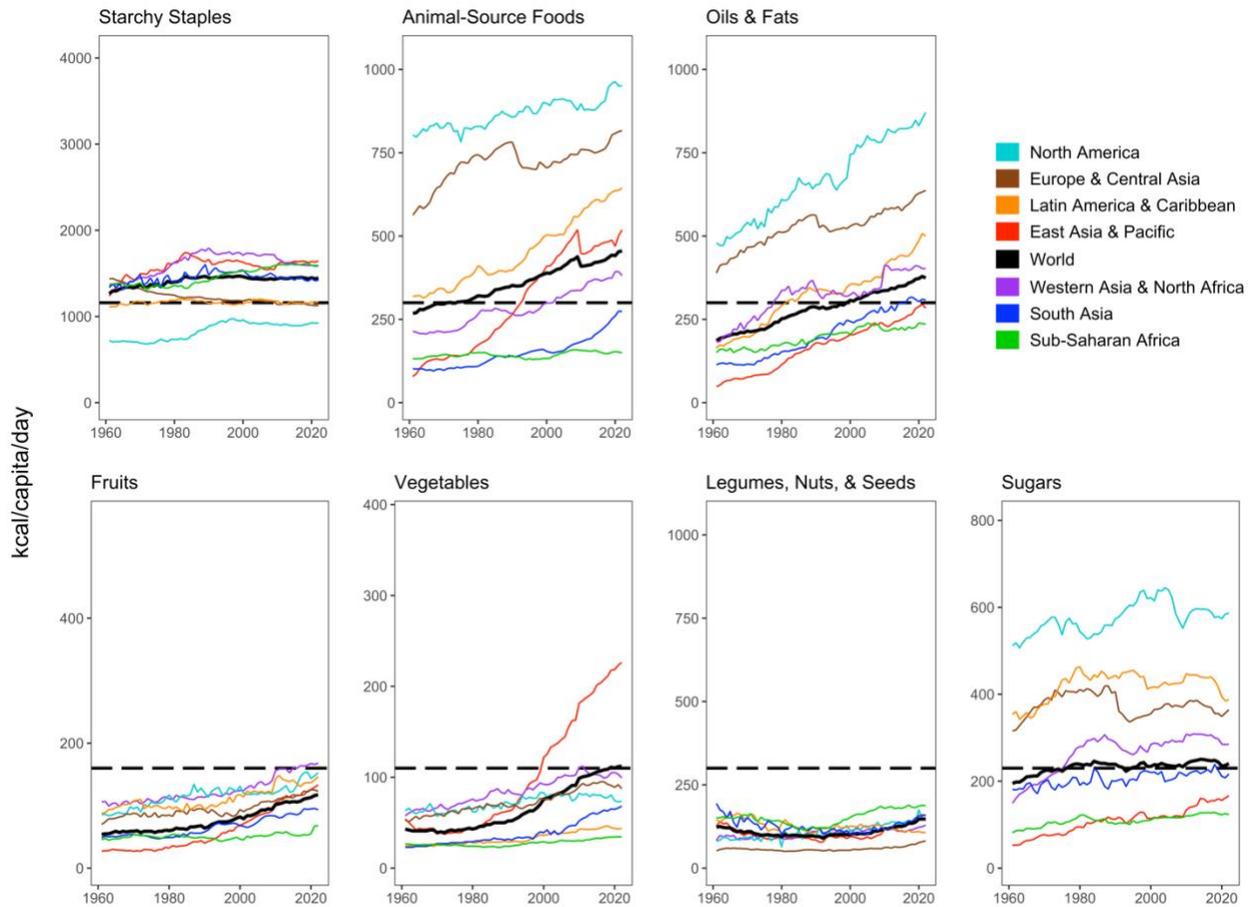

Note: Data shown are worldwide totals for daily per capita kilocalories available as food, computed by the authors from national data in FAOSTAT. Plots are scaled to reference intake values. Black dashed lines indicate Healthy Diet Basket targets used to measure access to a healthy diet as recommended in national dietary guidelines. Sugar availability is shown in reference to the World Health Organization guideline that sugar intake be limited to less than 10% of total dietary energy (WHO 2015).

Food groups and regions differ in both levels and changes over time, often with sustained plateaus or periods of either decline or rapid increase. Notable declines in availability include starchy staples in East Asia and the Pacific since their peak in the early 1980s, and in Europe and Central Asia, where supplies declined through the 1960s and 1970s before stabilizing near the HDB target. Supplies of legumes, nuts and seeds are below the HDB target in all regions and declined further in several regions before rising in Sub-Saharan Africa since the 1990s.

Availability of sugars rose in the 1960s and 1970s and then declined or plateaued in all regions except North America, where it continued to rise through the 1990s, and East Asia and the Pacific, where sugar continues to rise but remains well below the WHO's recommended limit.

Shortfalls in supply of healthy food groups below the HDB targets remain severe in several regions. For Sub-Saharan Africa, only starchy staples are available in adequate quantities, and only vegetable oils as well as legumes, nuts and seeds have increased significantly towards their HDB target. In South Asia, quantities of vegetables and fruits have increased, and oils and fats recently reached the HDB target in the late 2010s; while quantities, of sugars have stayed near the WHO limit since the 1980s, starchy staples have plateaued since the 1990s, and the legumes, nuts and seeds group declined from the world's highest levels in 1961 down to the global average and have barely risen since 2000. East Asia and the Pacific has had rapid growth in use of all food groups except starchy staples, and legumes, nuts and seeds, with availability rising above HDB targets since 1990 for animal source foods and since 2000 for vegetables, driven by extremely fast growth in vegetable availability in mainland China since the 1980s (see Supplementary Figure S2). In North America, starchy staples have remained constant since the 1990s at a level slightly below the target intake, while animal source foods, oils and fats, and sugars have remained at levels two or even three times higher than the target intake, suggesting that increased consumption of these food groups has more than compensated for this small deficit in energy intake from starchy staples.

To describe results over all food groups, we compute for each country the average percentage deficit over all food groups whose supply is at or below its HDB target. This computation produces a Healthy Diet Basket Index (HDBI), which summarizes the level of adequacy with respect to a diet needed for long-term health (HDBI=1 for complete adequacy across all six food groups). Figure 3 reports the mean level of adequacy over all countries in each region for each decade. For example, national food supplies in South Asia had an average HDBI of 0.45 in the 1960s, indicating that only 45% of HDB benchmarks were met. In the 2010s, this number rose to 0.65, representing an improvement of 20 percentage points towards food supply alignment with dietary guidelines.

**Figure 3. HDB index scores for adequacy of national food supplies by region and decade, 1960s-2020s**

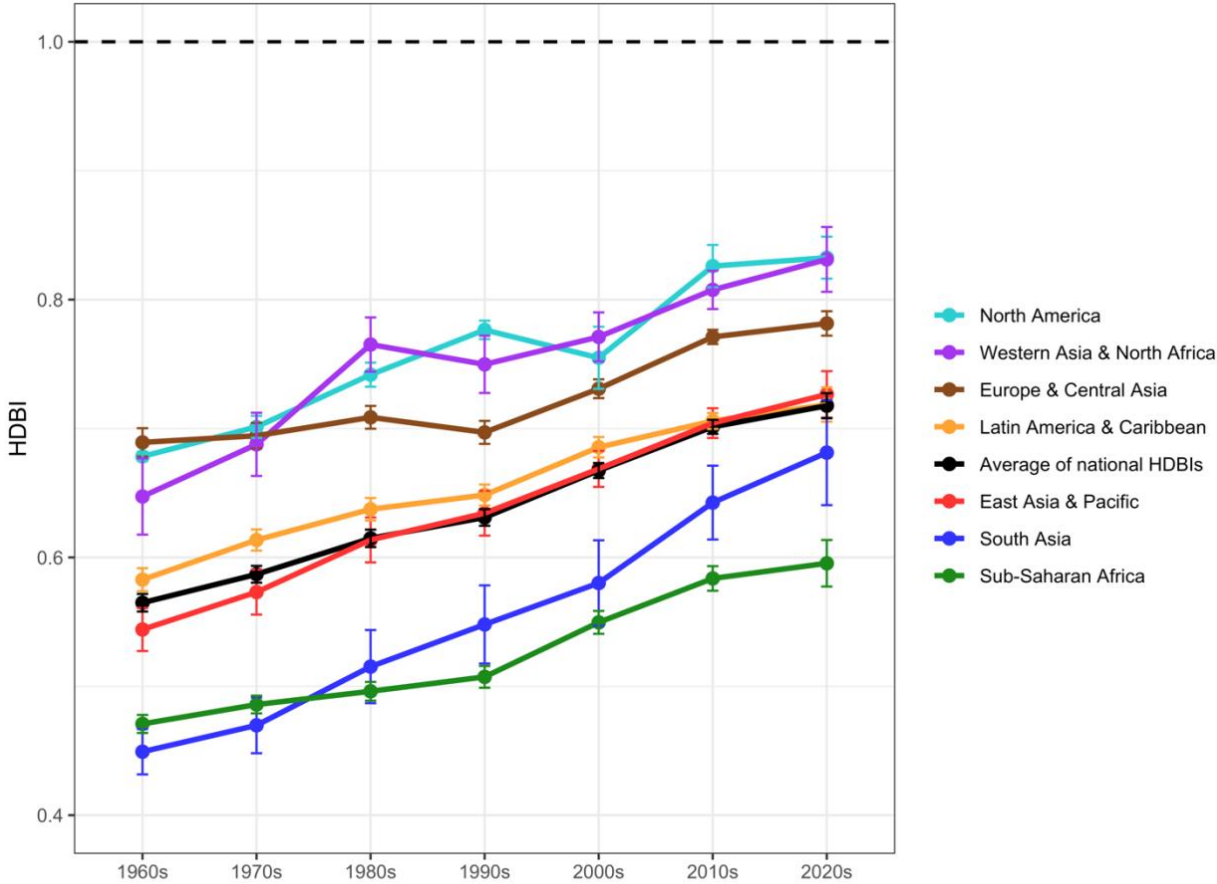

Note: Points show means of all countries' average proportion of target availability for the six Healthy Diet Basket (HDB) food groups, by geographic region and decade. Error bars indicate 95% confidence intervals. The black dashed line indicates the reference level where all food groups are supplied at or above the HDB target. The decade labelled as "2020s" includes data from 2020 through 2022.

The data in Figure 3 show how shortfalls below HDB targets are gradually closing over time, with faster progress in some regions than others. Average shortfalls across the food groups worsened during the 1980s in two regions (Europe and Central Asia, as well as Western Asia and North Africa) and worsened in the 2000s in North America, but otherwise have improved over time, especially in East Asia since the 1960s, South Asia since the 1970s, and Sub-Saharan Africa since the 1990s.

*Modeled projections*

Trends for adequacy of food supply will be affected by strong socioeconomic pressures (e.g., population and income growth) and by climate change, but they will also respond to policy changes. Figure 4 shows how future food adequacy could change between 2010 and 2050, using model simulations extracted from the IMPACT model (Sulser et al. 2021). Figure 4 shows projections based on a business-as-usual scenario and one scenario of increased investments in a limited set of commodities and improved water use.

**Figure 4. Projected HDB index scores for adequacy of national food supplies by region, 2010-2050**

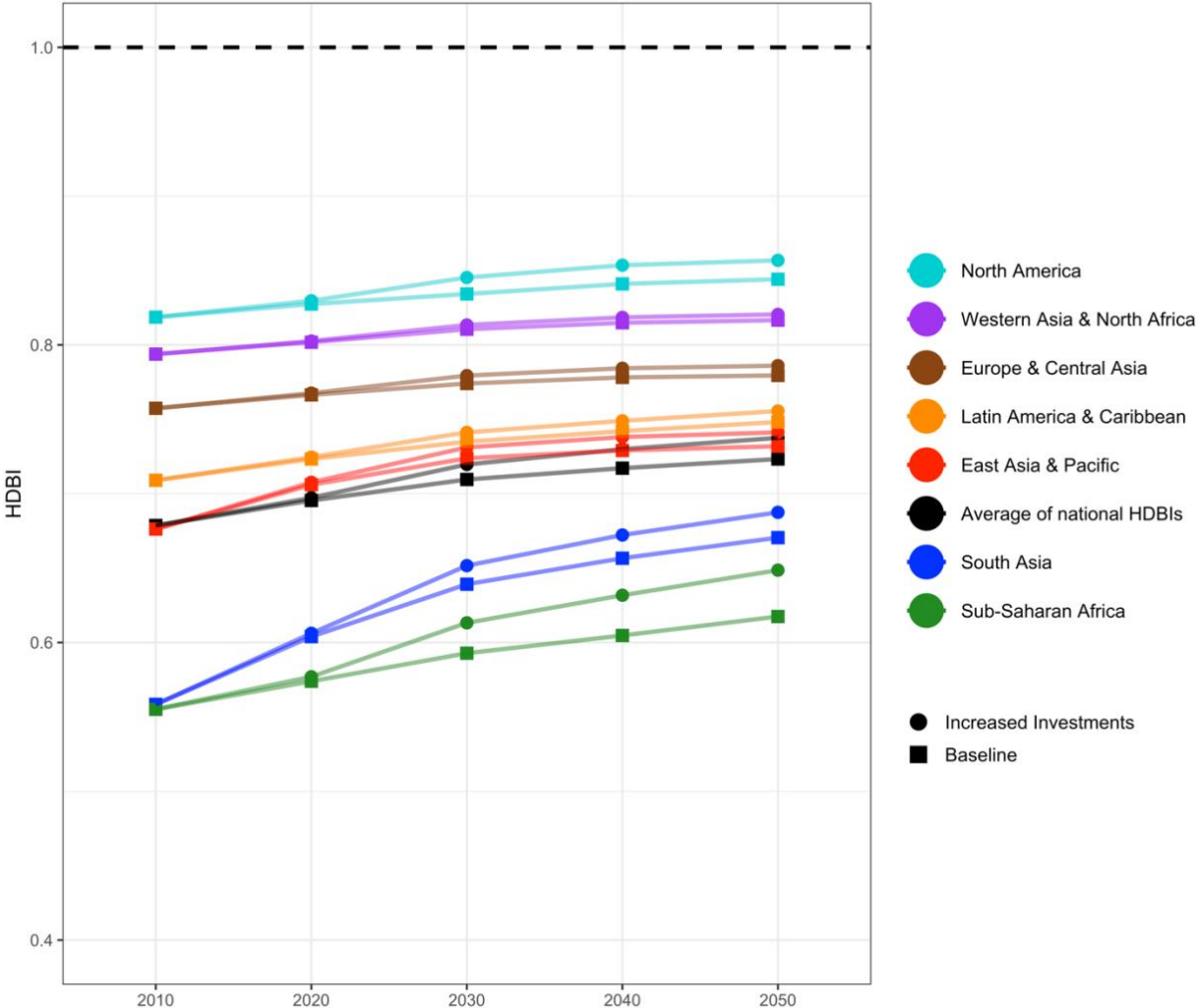

Note: Points show means of all countries' average proportion of target availability for the six Healthy Diet Basket (HDB) food groups, by geographic region and decade. Solid squares show baseline investment levels and solid circles show outcomes from an increased investments scenario. All outcomes are IMPACT model estimates given the SSP3 shared socioeconomic pathway and RCP8.5 climate-change scenario, using IMPACT to compute annual percentage changes in daily energy per capita of each commodity available for food use from the base year of observed quantities in 2010. Country inclusion described in the methods section and in Supplementary Table 5.

The IMPACT model's baseline for 2010 and 2020 differs somewhat from the average of countries' data reported to FAOSTAT shown in Figure 3 for the 2010s, due to differences in food group coverage and how regions are defined. From then onwards, model projections show faster improvement in regions with lower HDBI levels and slower improvements elsewhere.

The investment scenario represents supply-side enhancements that increase supply of some food groups (starchy staples, animal-source foods, pulses and oilseeds) and increase resource use efficiency and hence reduce consumer prices for those food groups. Within the increased investment scenario, the largest effect comes from the broad category of international R&D that leads to genetic improvement with accompanying changes in agronomic or veterinary management. This category of innovation has by far the most leverage, raising the productivity of all land, labor and other inputs used for each product, whereas the other forms of investment address important but much more limited opportunities for cost reduction or resource use. The availability of foods needed for healthy diets rises most quickly in South Asia and more slowly in Sub-Saharan Africa. In both of those regions, higher international R&D investments would accelerate progress by about a decade's worth of otherwise gradual change.

4**. Discussion and conclusion**

To meet global targets, Sub-Saharan Africa needs greater availability of all food groups except starchy staples; South Asia needs greater availability of vegetables, fruit, and animal source foods; East Asia and the Pacific need more fruit; and all regions of the world needs more legumes, nuts and seeds. Regional averages also obscure cases where disparities exist between or within countries; for example, East Asia and the Pacific is sufficient for vegetables only due to the outsize role of China, while nearby countries still face deep deficits. At current retail prices, even the least expensive locally available items in each HDB food group cost more per day than the available income for nearly 3 billion people worldwide (FAO, IFAD, UNICEF, WFP and WHO, 2024; World Bank, 2024). Investments in agriculture, including R&D and improved resource use, helps close the gaps faster than "business as usual." Despite both worsening environmental conditions due to climate change and relatively pessimistic socioeconomic conditions under SSP3 assumptions, these investments would lead to continued improvements, especially in South Asia and Sub-Saharan Africa. Our projections, however, show that investing in supply-side measures to lower prices for some food groups would bring food supplies only part way toward the quantities needed for healthy diets.

There are many factors that have driven food system advancements around the world since 1961, and that simultaneously explain some of the shortfalls that persist for key nutritious food groups, especially in Sub-Saharan Africa and South Asia. Government investments in agriculture commonly known as the Green Revolution brought innovations in crop and livestock genetics, irrigation and fertilizer technologies, cultivation equipment, and other aspects of food production systems. These innovations boosted productivity and helped to expand the availability of staple grains for human and animal consumption. Higher yields and the other labor-saving effects of new agricultural technologies led to higher incomes among farming households, which led to greater demand for aspirational foods such as meat, dairy, oils, fruits, and vegetables. However, these types of agricultural investments have never been fully implemented in Sub-Saharan Africa, and as a result the productivity growth that occurred in Asia and Latin America has not yet developed in the same way (Diao et al., 2008).

Each country and region of the world has its own story of food system development and transformation. In East Asia & Pacific, much of the exponential growth in nutrient-rich fruits, vegetables, and animal-source foods is driven by China, where the reforms of the 1970s and 1980s led to unprecedented economic growth. However, the role of income growth is a common theme across all countries and cultural histories. This analysis confirms Bennet's Law, the principle that demand for starchy staples declines as income rises. While per capita global and regional supplies for starchy staples have remained essentially unchanged since the 1960s, consumer demand for meat, milk, vegetable oils, vegetables, and fruits has risen substantially. The inequitable distribution of income growth around the world parallels the inequitable improvement in food supply adequacy that has meant availability of nutritious foods remains far below recommended targets in Sub-Saharan Africa. This pattern also suggests that ongoing income growth around the world will lead to continued expansion in the availability of all six recommended food groups. This expansion will only be equitably distributed if income growth is also equitable.

This analysis is subject to several limitations. There are well-known limitations to using FBS data, which have been extensively documented elsewhere ( Thar et al., 2020; Cafiero, 2013). Chief limitations include variation and error in country-level reporting, and the exclusion of foods from informal trade, foraging, and other unofficial sources, which may result in underreporting of totals for some foods. A change in the FAO methodology for balancing food supply data across input categories results in a discontinuity between data before and after 2010. We perform no adjustments to account for this discontinuity but instead present all data as published in FAOSTAT. Also, our approach concerns only the availability of raw ingredients and allows for the analysis of food system trends over time, as shifts in domestic production, trade policy, and research and development lead to increased or decreased availability of each food group. We do not account for consumer-level food waste; instead, our results reflect the potential availability of human food, a portion of which consumers may choose to discard or feed to pets. The FBS does not address whether primary ingredients that belong to the required HDB food groups are transformed into unhealthy foods that are not required in dietary guidelines through combination with additional ingredients, processing or contamination. Future work could expand the HDBI approach to account for excess energy intake and consumption of unhealthy foods. Yearly national data do not allow for estimates of subnational or temporal variation in food availability. This obscures the important observation that even in places where national food supplies are in full alignment with dietary guidelines, populations in given places and at given times may not in fact have physical or economic access to healthy diets.

We also do not account for inequities in access or distribution. This paper does not seek to assess access, which is discussed at length in the SOFI reports, but seeks to unpack one of its upstream determinants: availability. Food supplies provide information on the theoretical ability of all people to access adequate food given sufficient economic resources.

Does the world produce enough food? In calories, yes. In all foods needed for a healthy diet, no. Past improvements and projected future investments have closed some but not all the global shortfall in supply and demand of all food groups needed for a healthy diet, leaving large gaps for several food groups in specific regions of the world. Notably, past and projected investments leave significant shortfalls that are highly concentrated in specific food groups and geographic regions. Investments in "food" or "agriculture" need to go beyond the set of crops that have typically received the vast majority of investment – staple grains, starchy roots and tubers, and oil crops (Herforth et al., 2012, Pingali, 2015) – and focus on the foods that are the most in need: fruits, vegetables, legumes, nuts and seeds. Targeted investments would be needed to raise both demand and supply for locally-adapted foods.

These reconsidered supply side investments may also need to be accompanied by other policy levers to counteract income-driven trends toward greater consumption of animal-source foods and oils and fats. Although people with low-incomes are frequently unable to afford the foods required for a healthy diet, many high-income people choose to consume less-healthy diets even though healthy foods are available and affordable. Demand creation through higher incomes or social assistance as well as innovation may be needed to make under-consumed foods more attractive to consumers, alongside innovation to improve supply response, and open trade needed to keep prices low. Shifting consumption patterns toward healthy diets will require not only demand-side policy, but coordinated investment in R&D including a new focus on food groups for which there are large and persistent shortfalls in availability.

**Data availability:**

Historical food balance sheets data used in this study are publicly available from FAOSTAT as indicated in the References. IMPACT model results are available upon request.

**Code availability:**

All data analysis and visualizations were conducted in RStudio (version 2024.04.1+748). All R scripts will be available at https://sites.tufts.edu/foodpricesfornutrition/tools/.

**Supplementary Information**

**Figure S1. Projected HDBI scores showing adequacy of global and regional food supplies to meet healthy diet basket amounts, 2010-2050.**

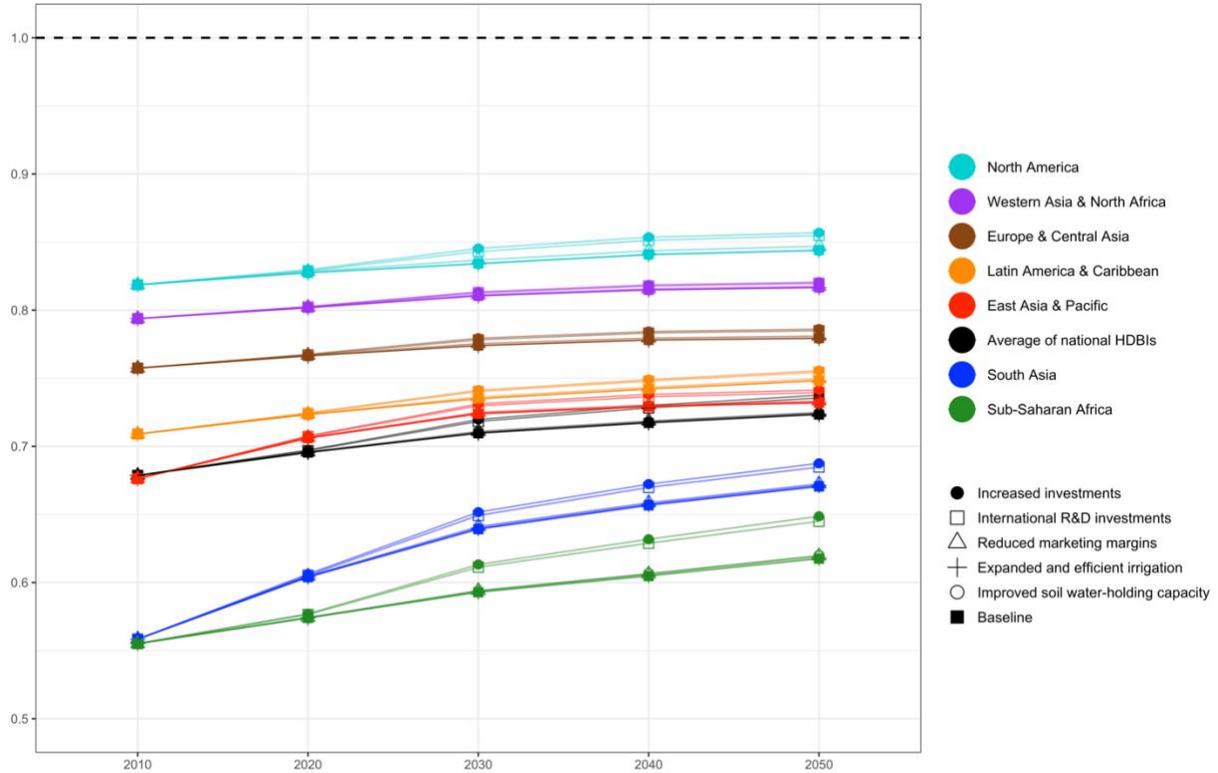

Note: Points show means of all countries' average proportion of target availability for the six HDB food groups, by geographic region and decade. Solid squares show business-as-usual investment levels, solid circles show outcomes from a higher level of investment in agricultural R&D complemented by improvements in water management and infrastructure, while the four other symbols shown in the legend show partial packages giving intermediate results. All outcomes are IMPACT model estimates given the SSP3 shared socioeconomic pathway and RCP8.5 climate-change scenario, using IMPACT to compute annual percentage changes in daily energy per capita of each commodity available for food use from the base year of observed quantities in 2010. Country inclusion described in the methods section and in Supplementary Table 5.

**Figure S2. Food supplies in East Asia & Pacific and mainland China relative to Healthy Diet Basket targets, 1961-2022.**

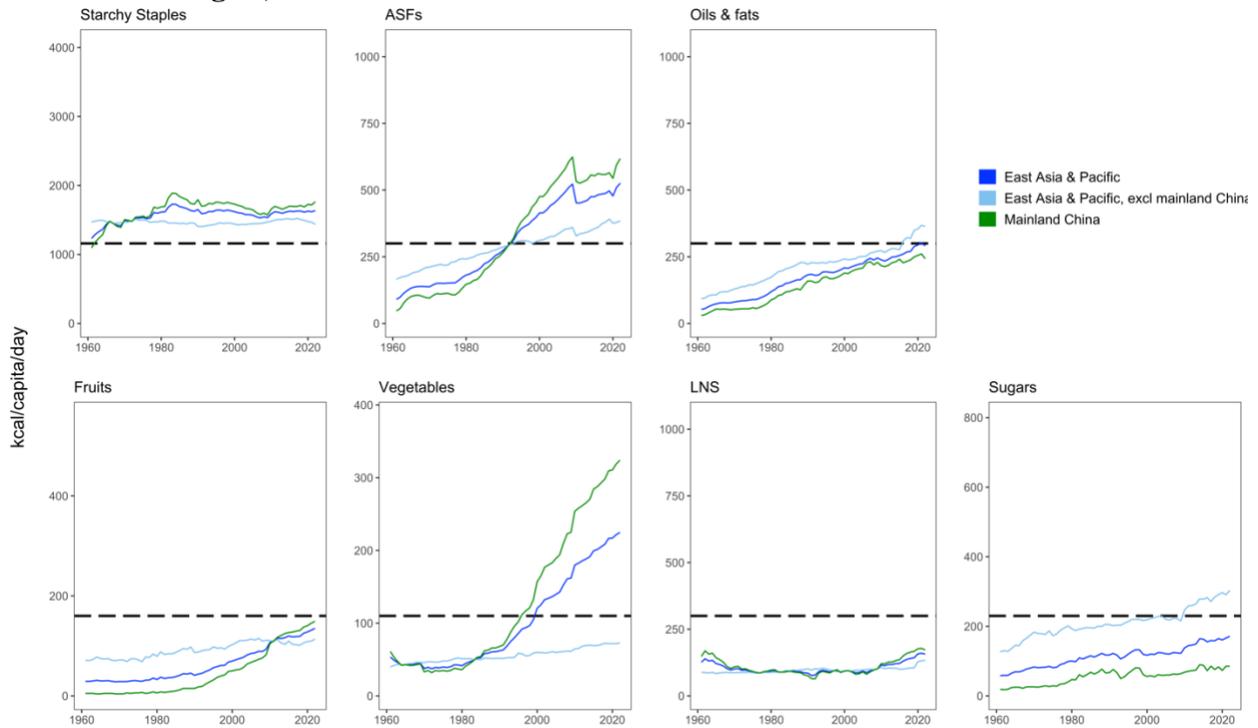

Note: Data shown are worldwide totals for daily per capita kilocalories available as food, computed by the authors from national data in FAOSTAT. Plots are scaled to reference intake values. Black dashed lines indicate Healthy Diet Basket targets used to measure access to a balanced diet as recommended in national dietary guidelines. Sugar availability is shown in reference to the World Health Organization guideline that sugar intake be limited to 10% or less of total daily energy (WHO 2015).

**Table S1. FAO commodity regroupings used in Figure 1.**

| FAO Commodity Code | FAO Commodity Name | New Grouping |
| --- | --- | --- |
| 2612 | Lemons, Limes and products | Other fruits |
| 2613 | Grapefruit and products | Other fruits |
| 2614 | Citrus, Other | Other fruits |
| 2618 | Pineapples and products | Other fruits |
| 2625 | Fruits, other | Other fruits |
| 2605 | Other vegetables | Other vegetables |
| 2775 | Aquatic plants | Other vegetables |
| 2768 | Meat, aquatic mammals | Other meats |
| 2736 | Edible offals | Other meats |
| 2735 | Meat, other | Other meats |
| 2516 | Oats | Other cereals |
| 2515 | Rye and products | Other cereals |
| 2513 | Barley and products | Other cereals |
| 2520 | Cereals, Other | Other cereals |
| 2549 | Other pulses and products | Other pulses and seeds |
| 2557 | Sunflower seeds | Other pulses and seeds |
| 2561 | Sesame seeds | Other pulses and seeds |
| 2535 | Yams | Other roots |
| 2534 | Roots, other | Other roots |
| 2769 | Aquatic Animals, Others | Other seafood |
| 2767 | Molluscs, Other | Other seafood |
| 2766 | Cephalopods | Other seafood |
| 2765 | Crustaceans | Other seafood |
| 2764 | Marine Fish, Other | Marine fish |
| 2762 | Demersal fish | Marine fish |
| 2763 | Pelagic fish | Marine fish |
| 2782 | Fish, Liver Oil | Other oils and oilcrops |
| 2781 | Fish, Body Oil | Other oils and oilcrops |
| 2579 | Sesameseed Oil | Other oils and oilcrops |
| 2581 | Ricebran Oil | Other oils and oilcrops |
| 2576 | Palmkernal Oil | Other oils and oilcrops |
| 2582 | Maize Germ Oil | Other oils and oilcrops |
| 2586 | Oilcrops Oil, Other | Other oils and oilcrops |
| 2562 | Palm kernels | Other oils and oilcrops |
| 2558 | Rape and mustardseed | Other oils and oilcrops |

| 2559 | Cottonseed | Other oils and oilcrops |
| 2743 | Cream | Other oils and oilcrops |
| 2570 | Oilcrops, Other | Other oils and oilcrops |
| 2543 | Sweeteners, Other | Other sweeteners |
| 2745 | Honey | Other sweeteners |
| 2536 | Sugar cane | Other sweeteners |
| 2541 | Non-centrifugal sugar | Other sweeteners |
| 2537 | Sugar beet | Other sweeteners |

Table S2. National HDBI and food supply as percentage of reference intake (first and last year available).

| Country | Year | Animal-Source Foods | Fruits | Legumes Nuts & Seeds | Oils & Fats | Starchy Staples | Sugars | Vegetables | HDBI |
|---|---|---|---|---|---|---|---|---|---|
| Afghanistan | 1961 | 0.663 | 0.369 | 0.117 | 0.243 | 2.203 | 0.222 | 0.2 | 0.432 |
| | 2022 | 0.421 | 0.352 | 0.21 | 0.822 | 1.395 | 0.37 | 0.364 | 0.528 |
| Albania | 1961 | 0.957 | 0.575 | 0.173 | 0.487 | 1.268 | 0.478 | 0.373 | 0.594 |
| | 2022 | 3.164 | 2.004 | 0.367 | 1.096 | 0.889 | 1.067 | 2.286 | 0.876 |
| Algeria | 1961 | 0.537 | 0.369 | 0.103 | 0.43 | 0.893 | 0.696 | 0.191 | 0.42 |
| | 2022 | 1.27 | 1.39 | 0.274 | 1.832 | 1.498 | 1.263 | 1.477 | 0.879 |
| Angola | 1961 | 0.31 | 0.412 | 0.337 | 0.41 | 1.083 | 0.37 | 0.164 | 0.439 |
| | 2022 | 0.58 | 0.828 | 0.358 | 0.862 | 1.334 | 0.417 | 0.331 | 0.66 |
| Antigua and Barbuda | 1961 | 1.153 | 0.75 | 0.047 | 0.837 | 0.779 | 1.535 | 0.055 | 0.578 |
| | 2022 | 2.224 | 1.325 | 0.132 | 1.284 | 0.589 | 0.934 | 0.5 | 0.704 |
| Argentina | 1961 | 2.987 | 0.531 | 0.093 | 0.813 | 1.047 | 1.661 | 0.509 | 0.658 |
| | 2022 | 3.054 | 0.543 | 0.379 | 1.815 | 0.972 | 1.614 | 0.513 | 0.735 |
| Armenia | 1992 | 1.087 | 0.525 | 0 | 0.13 | 1.213 | 0.948 | 0.691 | 0.724 |
| | 2022 | 2.662 | 1.023 | 0.236 | 1.43 | 1.068 | 1.332 | 1.359 | 0.873 |
| Australia | 1961 | 3.2 | 0.675 | 0.12 | 1.22 | 0.762 | 2.43 | 0.364 | 0.653 |
| | 2022 | 2.916 | 0.703 | 0.609 | 2.434 | 0.658 | 1.838 | 0.698 | 0.778 |
| Austria | 1961 | 2.253 | 1.1 | 0.09 | 1.81 | 0.978 | 1.704 | 0.373 | 0.74 |
| | 2022 | 2.654 | 0.909 | 0.428 | 3.515 | 0.849 | 1.661 | 0.738 | 0.821 |
| Azerbaijan | 1992 | 0.967 | 0.731 | 0.08 | 0.32 | 1.343 | 0.578 | 0.282 | 0.563 |
| | 2022 | 1.799 | 0.824 | 0.271 | 1.408 | 1.37 | 1.277 | 1.164 | 0.849 |
| Bahamas | 1961 | 1.673 | 0.606 | 0.163 | 0.72 | 0.825 | 1.583 | 0.864 | 0.696 |
| | 2022 | 2.586 | 1.418 | 0.154 | 1.149 | 0.472 | 1.323 | 0.892 | 0.753 |
| Bahrain | 2019 | 2.288 | 1.048 | 0.449 | 1.859 | 1.019 | 2.143 | 0.947 | 0.899 |
| | 2022 | 2.124 | 1.056 | 0.486 | 2.06 | 1.023 | 2.09 | 0.961 | 0.908 |
| Bangladesh | 1961 | 0.183 | 0.181 | 0.17 | 0.223 | 1.571 | 0.383 | 0.109 | 0.311 |
| | 2022 | 0.737 | 0.268 | 0.254 | 0.647 | 1.61 | 0.339 | 0.433 | 0.556 |
| Barbados | 1961 | 1.633 | 0.381 | 0.3 | 0.553 | 0.966 | 2.204 | 0.1 | 0.55 |
| | 2022 | 2.265 | 0.8 | 0.414 | 1.312 | 0.86 | 2.35 | 0.661 | 0.789 |
| Belarus | 1992 | 2.513 | 0.294 | 0.017 | 0.947 | 1.276 | 1.461 | 0.4 | 0.61 |
| | 2022 | 2.652 | 0.596 | 0.154 | 1.854 | 0.94 | 1.564 | 1.033 | 0.782 |
| Belgium | 2000 | 2.497 | 0.581 | 0.24 | 3.15 | 0.863 | 2.27 | 1.064 | 0.781 |
| | 2022 | 3.103 | 0.758 | 0.329 | 3.659 | 0.752 | 2.168 | 0.978 | 0.803 |
| Belgium-Luxembourg | 1961 | 1.96 | 0.5 | 0.1 | 2.193 | 0.9 | 1.161 | 0.518 | 0.67 |
| | 1999 | 2.473 | 0.819 | 0.193 | 3.143 | 0.826 | 2.139 | 1.2 | 0.806 |
| Belize | 1961 | 1.28 | 0.662 | 0.303 | 0.84 | 0.822 | 1.13 | 0.118 | 0.624 |
| | 2022 | 1.389 | 0.176 | 0.684 | 1.632 | 0.925 | 1.544 | 0.288 | 0.679 |

| Country | Year | | | | | | | | |
|---|---|---|---|---|---|---|---|---|---|
| Benin | 1961 | 0.207 | 0.369 | 0.367 | 0.527 | 1.122 | 0.104 | 0.1 | 0.428 |
| | 2022 | 0.446 | 0.268 | 0.89 | 1.057 | 1.489 | 0.381 | 0.408 | 0.669 |
| Bermuda | 1961 | 2.72 | 0.775 | 0.127 | 1.297 | 0.525 | 1.73 | 0.627 | 0.676 |
| | 2009 | 2.557 | 0.525 | 0.137 | 1.347 | 0.57 | 1.526 | 0.709 | 0.657 |
| Bhutan | 2019 | 0.833 | 0.225 | 0.173 | 1.582 | 1.767 | 0.645 | 0.412 | 0.607 |
| | 2022 | 0.775 | 0.281 | 0.229 | 1.494 | 1.8 | 0.634 | 0.396 | 0.614 |
| Bolivia | 1961 | 0.617 | 0.475 | 0.113 | 0.473 | 0.893 | 0.848 | 0.464 | 0.506 |
| | 2022 | 1.507 | 0.427 | 0.305 | 0.51 | 1.024 | 1.163 | 0.342 | 0.597 |
| Bosnia and Herzegovina | 1992 | 0.893 | 0.188 | 0.197 | 0.127 | 1.323 | 0.422 | 0.736 | 0.523 |
| | 2022 | 1.988 | 1.554 | 0.362 | 0.88 | 1.145 | 1.258 | 1.53 | 0.874 |
| Botswana | 1961 | 0.917 | 0.081 | 0.77 | 0.26 | 0.975 | 0.717 | 0.091 | 0.516 |
| | 2022 | 1.03 | 0.298 | 0.158 | 0.977 | 1.143 | 1.132 | 0.269 | 0.617 |
| Brazil | 1961 | 0.793 | 0.631 | 0.773 | 0.46 | 0.9 | 1.783 | 0.155 | 0.619 |
| | 2022 | 2.62 | 0.97 | 0.405 | 2.326 | 0.881 | 1.648 | 0.335 | 0.765 |
| Brunei Darussalam | 1961 | 0.743 | 0.35 | 0.167 | 0.53 | 0.922 | 1.474 | 0.191 | 0.484 |
| | 2009 | 1.79 | 0.544 | 0.283 | 0.977 | 1.141 | 1.691 | 0.564 | 0.728 |
| Bulgaria | 1961 | 1.12 | 0.738 | 0.223 | 1.133 | 1.653 | 0.861 | 0.573 | 0.756 |
| | 2022 | 2.031 | 0.556 | 0.314 | 2.265 | 0.888 | 1.39 | 0.592 | 0.725 |
| Burkina Faso | 1961 | 0.21 | 0.038 | 0.813 | 0.1 | 0.701 | 0.074 | 0.136 | 0.333 |
| | 2022 | 0.689 | 0.439 | 1.843 | 0.352 | 1.332 | 0.238 | 0.389 | 0.645 |
| Burundi | 2010 | 0.098 | 1.698 | 0.765 | 0.212 | 0.668 | 0.14 | 0.279 | 0.504 |
| | 2022 | 0.122 | 0.822 | 0.894 | 0.223 | 0.967 | 0.265 | 0.236 | 0.544 |
| Cabo Verde | 1961 | 0.167 | 0.588 | 0.4 | 0.153 | 0.981 | 0.626 | 0.045 | 0.389 |
| | 2022 | 1.202 | 0.47 | 0.377 | 0.976 | 1.199 | 0.911 | 0.481 | 0.717 |
| Cambodia | 1961 | 0.173 | 0.344 | 0.127 | 0.083 | 1.37 | 0.465 | 0.3 | 0.338 |
| | 2022 | 0.616 | 0.119 | 0.193 | 0.483 | 1.784 | 0.425 | 0.234 | 0.441 |
| Cameroon | 1961 | 0.29 | 0.1 | 0.53 | 0.26 | 1.338 | 0.096 | 0.218 | 0.4 |
| | 2022 | 0.457 | 0.38 | 0.882 | 0.908 | 1.554 | 0.583 | 0.647 | 0.712 |
| Canada | 1961 | 2.45 | 0.637 | 0.213 | 1.65 | 0.654 | 2.048 | 0.536 | 0.674 |
| | 2022 | 2.332 | 0.864 | 0.665 | 3.02 | 0.856 | 1.887 | 0.751 | 0.856 |
| Central African Republic | 1961 | 0.25 | 0.3 | 0.143 | 0.96 | 1.499 | 0.117 | 0.1 | 0.459 |
| | 2022 | 0.736 | 0.423 | 1.412 | 1.178 | 0.847 | 0.331 | 0.132 | 0.69 |
| Chad | 1961 | 0.487 | 0.288 | 1.21 | 0.407 | 1.31 | 0.213 | 0.073 | 0.542 |
| | 2022 | 1.157 | 0.082 | 1.041 | 0.569 | 1.06 | 0.422 | 0.04 | 0.615 |
| Chile | 1961 | 1.14 | 0.306 | 0.19 | 0.683 | 1.167 | 1.165 | 0.782 | 0.66 |
| | 2022 | 2.129 | 0.481 | 0.231 | 1.196 | 1.046 | 1.998 | 0.604 | 0.719 |
| China, Hong Kong SAR | 1961 | 1.087 | 0.238 | 0.507 | 1.18 | 1.178 | 0.891 | 1.082 | 0.791 |

| Country | Year | | | | | | | | |
|---|---|---|---|---|---|---|---|---|---|
| | 2022 | 2.908 | 0.622 | 0.361 | 1.415 | 0.708 | 1.204 | 1.377 | 0.782 |
| China, Macao SAR | 1961 | 0.767 | 0.162 | 0.443 | 1 | 1.438 | 1 | 0.418 | 0.632 |
| | 2022 | 2.501 | 0.507 | 0.242 | 1.617 | 0.791 | 1.071 | 0.884 | 0.737 |
| China, mainland | 1961 | 0.157 | 0.031 | 0.493 | 0.1 | 0.945 | 0.083 | 0.555 | 0.38 |
| | 2022 | 2.06 | 0.932 | 0.575 | 0.81 | 1.521 | 0.372 | 2.946 | 0.886 |
| China, Taiwan Province of | 1961 | 0.703 | 0.15 | 0.35 | 0.33 | 1.547 | 0.622 | 0.555 | 0.515 |
| | 2022 | 2.034 | 0.823 | 0.616 | 2.272 | 0.759 | 1.118 | 0.934 | 0.855 |
| Colombia | 1961 | 1.073 | 0.244 | 0.16 | 0.387 | 0.882 | 2.452 | 0.155 | 0.471 |
| | 2022 | 1.942 | 1.093 | 0.22 | 1.433 | 0.971 | 2.573 | 0.363 | 0.759 |
| Comoros | 2010 | 0.558 | 1.788 | 0.637 | 1.009 | 1.145 | 0.465 | 0.093 | 0.715 |
| | 2022 | 0.619 | 2.315 | 0.617 | 0.428 | 1.157 | 0.634 | 0.198 | 0.644 |
| Congo | 1961 | 0.357 | 0.238 | 0.33 | 0.563 | 1.384 | 0.083 | 0.082 | 0.428 |
| | 2022 | 0.914 | 0.294 | 0.164 | 0.937 | 1.107 | 0.614 | 0.118 | 0.571 |
| Costa Rica | 1961 | 0.783 | 0.6 | 0.31 | 0.703 | 0.734 | 1.896 | 0.109 | 0.54 |
| | 2022 | 2.255 | 0.615 | 0.408 | 1.634 | 0.799 | 1.944 | 0.341 | 0.694 |
| Cote d'Ivoire | 1961 | 0.423 | 0.075 | 0.5 | 0.37 | 1.523 | 0.343 | 0.273 | 0.44 |
| | 2022 | 0.434 | 0.079 | 0.245 | 0.929 | 1.883 | 0.546 | 0.302 | 0.498 |
| Croatia | 1992 | 1.543 | 0.481 | 0.12 | 0.96 | 0.728 | 1.704 | 0.355 | 0.607 |
| | 2022 | 2.638 | 0.737 | 0.235 | 2.04 | 0.767 | 1.861 | 1.371 | 0.79 |
| Cuba | 1961 | 1 | 0.244 | 0.473 | 0.807 | 0.74 | 2.074 | 0.209 | 0.579 |
| | 2022 | 1.559 | 0.667 | 0.447 | 0.988 | 1.356 | 2.06 | 0.754 | 0.809 |
| Cyprus | 1961 | 0.967 | 1.225 | 0.64 | 1.22 | 0.923 | 0.948 | 0.5 | 0.838 |
| | 2022 | 2.411 | 0.587 | 0.327 | 1.48 | 0.949 | 1.108 | 0.625 | 0.748 |
| Czechia | 1993 | 2.49 | 0.506 | 0.13 | 1.893 | 0.772 | 1.687 | 0.427 | 0.639 |
| | 2022 | 2.4 | 0.639 | 0.228 | 3.122 | 0.676 | 1.544 | 0.562 | 0.684 |
| Czechoslovakia | 1961 | 2.03 | 0.419 | 0.093 | 1.717 | 1.214 | 1.922 | 0.4 | 0.652 |
| | 1992 | 2.19 | 0.469 | 0.093 | 1.89 | 0.928 | 1.965 | 0.364 | 0.642 |
| Democratic Republic of the Congo | 2010 | 0.114 | 0.141 | 0.273 | 0.444 | 1.637 | 0.156 | 0.061 | 0.339 |
| | 2022 | 0.119 | 0.153 | 0.23 | 0.358 | 1.601 | 0.149 | 0.045 | 0.317 |
| Denmark | 1961 | 2.08 | 0.469 | 0.023 | 2.667 | 0.872 | 2.165 | 0.245 | 0.602 |
| | 2022 | 3.191 | 0.789 | 0.265 | 1.928 | 0.835 | 2.154 | 0.693 | 0.764 |
| Djibouti | 1961 | 0.697 | 0 | 0.093 | 0.283 | 0.765 | 1.517 | 0.055 | 0.482 |
| | 2022 | 0.548 | 0.299 | 0.689 | 1.247 | 1.261 | 1.573 | 0.453 | 0.665 |
| Dominica | 1961 | 0.733 | 1.1 | 0.043 | 0.423 | 0.624 | 1.509 | 0.336 | 0.527 |
| | 2022 | 2.282 | 2.644 | 0.2 | 0.737 | 0.74 | 1.172 | 0.817 | 0.749 |

| Country | Year | | | | | | | | |
|---|---|---|---|---|---|---|---|---|---|
| Dominican Republic | 1961 | 0.607 | 1.225 | 0.387 | 0.387 | 0.72 | 0.987 | 0.109 | 0.535 |
| | 2022 | 1.71 | 3.236 | 0.292 | 1.492 | 0.891 | 1.817 | 0.821 | 0.834 |
| Ecuador | 1961 | 0.93 | 1.425 | 0.417 | 0.637 | 0.745 | 1.117 | 0.355 | 0.68 |
| | 2022 | 1.553 | 0.263 | 0.111 | 1.951 | 0.902 | 1.236 | 0.17 | 0.574 |
| Egypt | 1961 | 0.337 | 0.456 | 0.267 | 0.523 | 1.247 | 0.6 | 0.655 | 0.54 |
| | 2022 | 0.944 | 1.033 | 0.145 | 0.704 | 1.742 | 1.108 | 0.859 | 0.775 |
| El Salvador | 1961 | 0.57 | 0.288 | 0.23 | 0.417 | 0.785 | 0.909 | 0.155 | 0.407 |
| | 2022 | 1.415 | 0.642 | 0.581 | 0.834 | 1.198 | 1.441 | 0.549 | 0.768 |
| Estonia | 1992 | 3.067 | 0.2 | 0.01 | 0.58 | 0.888 | 0.817 | 0.318 | 0.499 |
| | 2022 | 3.316 | 0.734 | 0.309 | 1.877 | 0.618 | 1.58 | 0.593 | 0.709 |
| Eswatini | 1961 | 0.94 | 0.181 | 0.433 | 0.187 | 1.16 | 1.209 | 0.082 | 0.471 |
| | 2022 | 0.809 | 0.483 | 0.206 | 0.787 | 1.224 | 1.437 | 0.197 | 0.58 |
| Ethiopia | 1993 | 0.227 | 0.025 | 0.317 | 0.143 | 1.06 | 0.157 | 0.064 | 0.296 |
| | 2022 | 0.339 | 0.143 | 0.783 | 0.619 | 1.534 | 0.336 | 0.123 | 0.501 |
| Ethiopia PDR | 1961 | 0.487 | 0.019 | 0.603 | 0.24 | 1.173 | 0.096 | 0.055 | 0.401 |
| | 1992 | 0.273 | 0.025 | 0.49 | 0.273 | 1.012 | 0.13 | 0.055 | 0.353 |
| Fiji | 1961 | 0.623 | 1.556 | 0.123 | 0.523 | 1.409 | 1.235 | 0.091 | 0.56 |
| | 2022 | 1.275 | 0.769 | 0.29 | 1.412 | 1.118 | 1.848 | 0.731 | 0.798 |
| Finland | 1961 | 3.033 | 0.356 | 0.047 | 1.567 | 1.105 | 1.913 | 0.127 | 0.588 |
| | 2022 | 3.399 | 0.645 | 0.293 | 1.233 | 0.991 | 1.449 | 0.606 | 0.756 |
| France | 1961 | 2.737 | 0.45 | 0.147 | 1.263 | 1.002 | 1.287 | 0.936 | 0.756 |
| | 2022 | 3.035 | 0.832 | 0.255 | 2.353 | 0.867 | 1.616 | 0.697 | 0.775 |
| French Polynesia | 1961 | 1.237 | 0.887 | 0.12 | 0.71 | 1.181 | 1.517 | 0.3 | 0.67 |
| | 2022 | 2.33 | 1.032 | 0.187 | 1.508 | 0.86 | 1.062 | 0.369 | 0.736 |
| Gabon | 1961 | 0.707 | 0.262 | 0.123 | 0.243 | 1.228 | 0.113 | 0.245 | 0.43 |
| | 2022 | 1.3 | 0.186 | 0.585 | 0.255 | 1.353 | 0.668 | 0.232 | 0.543 |
| Gambia | 1961 | 0.337 | 0.044 | 0.6 | 0.803 | 1.191 | 0.487 | 0.073 | 0.476 |
| | 2022 | 0.62 | 0.041 | 0.423 | 0.899 | 1.1 | 2.288 | 0.162 | 0.524 |
| Georgia | 1992 | 0.893 | 0.594 | 0.11 | 0.18 | 1.068 | 0.5 | 0.255 | 0.505 |
| | 2022 | 1.912 | 0.577 | 0.167 | 0.988 | 1.274 | 1.676 | 0.435 | 0.695 |
| Germany | 1961 | 1.927 | 0.681 | 0.123 | 2.003 | 0.835 | 1.465 | 0.282 | 0.654 |
| | 2022 | 2.916 | 0.804 | 0.353 | 2.893 | 0.733 | 1.987 | 0.681 | 0.762 |
| Ghana | 1961 | 0.353 | 0.537 | 0.147 | 0.417 | 1.302 | 0.435 | 0.173 | 0.438 |
| | 2022 | 0.494 | 0.354 | 0.475 | 0.699 | 2.011 | 0.539 | 0.194 | 0.536 |
| Greece | 1961 | 1.123 | 1.25 | 0.423 | 1.527 | 1.196 | 0.648 | 0.618 | 0.84 |
| | 2022 | 2.668 | 1.338 | 0.514 | 2.235 | 0.775 | 1.497 | 0.737 | 0.838 |
| Grenada | 1961 | 0.633 | 1.819 | 0.237 | 0.527 | 0.58 | 1.187 | 0.155 | 0.522 |
| | 2022 | 2.21 | 1.245 | 0.294 | 1.184 | 0.487 | 1.363 | 0.465 | 0.708 |

| Country | Year | | | | | | | | |
|---|---|---|---|---|---|---|---|---|---|
| Guatemala | 1961 | 0.463 | 0.288 | 0.307 | 0.247 | 1.038 | 0.909 | 0.164 | 0.411 |
| | 2022 | 1.04 | 1.021 | 0.422 | 0.901 | 1.069 | 2.458 | 0.517 | 0.807 |
| Guinea | 1961 | 0.147 | 0.35 | 0.297 | 0.717 | 1.166 | 0.126 | 1.509 | 0.585 |
| | 2022 | 0.443 | 0.54 | 0.368 | 1.27 | 1.676 | 0.811 | 0.422 | 0.629 |
| Guinea-Bissau | 1961 | 0.403 | 0.144 | 0.26 | 0.93 | 0.92 | 0.057 | 0.1 | 0.459 |
| | 2022 | 0.415 | 0.196 | 0.266 | 1.277 | 1.359 | 0.112 | 0.131 | 0.501 |
| Guyana | 1961 | 0.837 | 0.537 | 0.207 | 0.677 | 1.078 | 1.696 | 0.091 | 0.558 |
| | 2022 | 1.953 | 1.31 | 0.278 | 0.564 | 1.231 | 1.453 | 1.765 | 0.807 |
| Haiti | 1961 | 0.273 | 0.675 | 0.47 | 0.19 | 0.978 | 1.1 | 0.218 | 0.467 |
| | 2022 | 0.415 | 0.815 | 0.276 | 0.579 | 0.983 | 0.963 | 0.125 | 0.532 |
| Honduras | 1961 | 0.56 | 0.713 | 0.377 | 0.25 | 0.974 | 1.078 | 0.082 | 0.493 |
| | 2022 | 1.018 | 0.284 | 0.387 | 1.732 | 0.985 | 1.782 | 0.226 | 0.647 |
| Hungary | 1961 | 1.827 | 0.519 | 0.1 | 1.293 | 1.155 | 1.278 | 0.527 | 0.691 |
| | 2022 | 2.197 | 0.626 | 0.091 | 3.157 | 0.814 | 1.918 | 0.614 | 0.691 |
| Iceland | 1961 | 4.473 | 0.394 | 0.05 | 1.393 | 0.674 | 2.435 | 0.073 | 0.532 |
| | 2022 | 4.02 | 0.852 | 0.179 | 1.633 | 0.918 | 1.51 | 0.594 | 0.757 |
| India | 1961 | 0.3 | 0.3 | 0.767 | 0.417 | 1.11 | 0.835 | 0.218 | 0.5 |
| | 2022 | 0.868 | 0.637 | 0.654 | 1.016 | 1.196 | 0.924 | 0.704 | 0.81 |
| Indonesia | 1961 | 0.167 | 0.488 | 0.273 | 0.263 | 1.206 | 0.504 | 0.164 | 0.392 |
| | 2022 | 0.863 | 0.928 | 0.522 | 1.366 | 1.276 | 1.333 | 0.419 | 0.789 |
| Iran | 1961 | 0.603 | 0.55 | 0.223 | 0.227 | 0.927 | 0.987 | 0.264 | 0.466 |
| | 2022 | 0.878 | 1.222 | 0.401 | 1.282 | 1.309 | 1.341 | 0.661 | 0.823 |
| Iraq | 1961 | 0.683 | 0.375 | 0.127 | 0.493 | 0.859 | 0.957 | 0.564 | 0.517 |
| | 2022 | 0.463 | 0.748 | 0.507 | 1.214 | 1.348 | 0.957 | 0.627 | 0.724 |
| Ireland | 1961 | 2.847 | 0.369 | 0.073 | 1.45 | 1.141 | 2.261 | 0.245 | 0.615 |
| | 2022 | 4.324 | 0.674 | 0.204 | 2.004 | 0.878 | 1.978 | 0.729 | 0.747 |
| Israel | 1961 | 1.463 | 1.087 | 0.327 | 1.157 | 1.084 | 1.496 | 0.855 | 0.864 |
| | 2022 | 3.034 | 1.315 | 0.76 | 2.201 | 1.047 | 1.617 | 1.249 | 0.96 |
| Italy | 1961 | 1.307 | 0.838 | 0.257 | 1.243 | 1.228 | 1.061 | 0.718 | 0.802 |
| | 2022 | 2.807 | 0.991 | 0.42 | 2.846 | 0.968 | 1.598 | 0.511 | 0.815 |
| Jamaica | 1961 | 0.763 | 1.225 | 0.083 | 0.62 | 0.728 | 1.674 | 0.109 | 0.551 |
| | 2022 | 1.69 | 0.782 | 0.147 | 1.277 | 1.038 | 1.8 | 0.744 | 0.779 |
| Japan | 1961 | 0.743 | 0.2 | 0.54 | 0.38 | 1.433 | 0.761 | 0.564 | 0.571 |
| | 2022 | 1.82 | 0.342 | 0.505 | 1.369 | 0.903 | 1.208 | 0.707 | 0.743 |
| Jordan | 1961 | 0.363 | 0.869 | 0.263 | 0.653 | 1.006 | 1.143 | 1.136 | 0.691 |
| | 2022 | 1.041 | 0.395 | 0.404 | 1.572 | 0.937 | 1.628 | 0.7 | 0.739 |
| Kazakhstan | 1992 | 2.183 | 0.112 | 0.02 | 0.827 | 1.472 | 0.943 | 0.3 | 0.543 |
| | 2022 | 3.128 | 0.766 | 0.532 | 1.844 | 0.896 | 1.357 | 1.523 | 0.866 |
| Kenya | 1961 | 0.783 | 0.637 | 0.817 | 0.14 | 1.212 | 0.609 | 0.164 | 0.59 |
| | 2022 | 0.745 | 0.67 | 0.589 | 0.699 | 0.917 | 1.091 | 0.391 | 0.669 |

| Country | Year | | | | | | | | |
|---|---|---|---|---|---|---|---|---|---|
| Kiribati | 1961 | 0.783 | 5.281 | 0.083 | 0.653 | 0.747 | 0.922 | 0.345 | 0.602 |
| | 2022 | 1.348 | 5.096 | 0.111 | 0.975 | 0.888 | 1.912 | 0.285 | 0.71 |
| Kuwait | 1961 | 1.57 | 0.644 | 0.14 | 1.093 | 0.924 | 1.565 | 0.836 | 0.757 |
| | 2022 | 2.082 | 1.58 | 0.344 | 1.62 | 1.232 | 1.475 | 0.987 | 0.889 |
| Kyrgyzstan | 1992 | 1.9 | 0.219 | 0.047 | 0.703 | 1.266 | 1.113 | 0.536 | 0.584 |
| | 2022 | 2.162 | 0.328 | 0.647 | 0.592 | 1.086 | 0.876 | 1.39 | 0.761 |
| Lao People's Democratic Republic | 1961 | 0.26 | 0.138 | 0.117 | 0.093 | 1.474 | 0.035 | 0.127 | 0.289 |
| | 2022 | 0.831 | 1.162 | 0.197 | 0.346 | 1.831 | 0.911 | 1.263 | 0.729 |
| Latvia | 1992 | 2.677 | 0.212 | 0 | 1.197 | 1.323 | 1.678 | 0.5 | 0.785 |
| | 2022 | 2.808 | 0.545 | 0.143 | 2.326 | 0.783 | 1.645 | 0.596 | 0.678 |
| Lebanon | 1961 | 0.837 | 1.138 | 0.46 | 0.883 | 1.082 | 0.909 | 0.627 | 0.801 |
| | 2022 | 1.232 | 0.783 | 0.547 | 1.4 | 1.016 | 2.05 | 0.91 | 0.873 |
| Lesotho | 1961 | 0.44 | 0.112 | 0.357 | 0.073 | 1.364 | 0.478 | 0.082 | 0.344 |
| | 2022 | 0.77 | 0.142 | 0.132 | 0.353 | 1.014 | 0.573 | 0.151 | 0.425 |
| Liberia | 1961 | 0.307 | 0.556 | 0.17 | 0.467 | 1.358 | 0.1 | 0.273 | 0.462 |
| | 2022 | 0.446 | 0.325 | 0.127 | 0.738 | 1.298 | 0.337 | 0.19 | 0.471 |
| Libya | 2010 | 1.208 | 1.046 | 0.219 | 1.724 | 1.206 | 1.541 | 1.089 | 0.87 |
| | 2022 | 1.555 | 1.177 | 0.48 | 1.708 | 1.073 | 1.525 | 1.183 | 0.913 |
| Lithuania | 1992 | 2.147 | 0.231 | 0.053 | 1.117 | 1.349 | 1.157 | 0.418 | 0.617 |
| | 2022 | 3.405 | 0.542 | 0.297 | 1.403 | 0.93 | 1.679 | 0.664 | 0.739 |
| Luxembourg | 2000 | 3.58 | 1.044 | 0.043 | 1.407 | 0.727 | 1.257 | 0.555 | 0.721 |
| | 2022 | 3.271 | 0.784 | 0.153 | 1.254 | 0.868 | 1.266 | 0.707 | 0.752 |
| Madagascar | 1961 | 0.897 | 0.488 | 0.253 | 0.173 | 1.547 | 0.396 | 0.191 | 0.5 |
| | 2022 | 0.228 | 0.356 | 0.062 | 0.264 | 1.375 | 0.606 | 0.092 | 0.334 |
| Malawi | 1961 | 0.137 | 0.331 | 1.19 | 0.14 | 1.252 | 0.113 | 0.164 | 0.462 |
| | 2022 | 0.715 | 1.181 | 0.521 | 0.39 | 1.53 | 0.477 | 0.511 | 0.689 |
| Malaysia | 1961 | 0.74 | 0.944 | 0.19 | 0.7 | 1.235 | 1.3 | 0.136 | 0.618 |
| | 2022 | 1.77 | 0.592 | 0.29 | 1.851 | 1.038 | 1.747 | 0.54 | 0.737 |
| Maldives | 1961 | 0.13 | 0.688 | 0.133 | 0.513 | 0.689 | 1.522 | 0.409 | 0.427 |
| | 2022 | 1.947 | 0.773 | 0.51 | 0.533 | 0.952 | 1.041 | 0.888 | 0.776 |
| Mali | 1961 | 0.623 | 0.119 | 0.283 | 0.273 | 1.051 | 0.113 | 0.155 | 0.409 |
| | 2022 | 0.442 | 0.698 | 0.431 | 0.546 | 1.75 | 0.384 | 0.771 | 0.648 |
| Malta | 1961 | 1.5 | 0.388 | 0.463 | 1.233 | 1.154 | 1.591 | 0.464 | 0.719 |
| | 2022 | 2.302 | 0.691 | 0.205 | 1.567 | 1.019 | 1.976 | 0.828 | 0.787 |
| Marshall Islands | 2019 | 1.694 | 3.652 | 0.064 | 1.082 | 0.987 | 0.711 | 0.742 | 0.799 |
| | 2022 | 2.396 | 3.659 | 0.086 | 1.504 | 0.881 | 0.901 | 0.685 | 0.775 |
| Mauritania | 1961 | 2.167 | 0.425 | 0.487 | 0.28 | 0.946 | 0.857 | 0.018 | 0.526 |

| Country | Year | | | | | | | | |
|---|---|---|---|---|---|---|---|---|---|
| | 2022 | 1.372 | 0.31 | 0.302 | 0.874 | 1.458 | 1.867 | 0.395 | 0.647 |
| Mauritius | 1961 | 0.523 | 0.094 | 0.307 | 1.05 | 1.133 | 1.809 | 0.164 | 0.515 |
| | 2022 | 1.69 | 0.397 | 0.405 | 1.879 | 1.124 | 1.493 | 0.589 | 0.732 |
| Mexico | 1961 | 0.843 | 0.438 | 0.623 | 0.547 | 1.147 | 1.061 | 0.118 | 0.595 |
| | 2022 | 2.27 | 1.037 | 0.379 | 1.451 | 1.177 | 1.746 | 0.432 | 0.802 |
| Micronesia (Federated States of) | 2019 | 1.457 | 2.888 | 0.028 | 1.239 | 0.95 | 1.045 | 0.214 | 0.699 |
| | 2022 | 1.43 | 3.655 | 0.05 | 1.16 | 0.891 | 1.072 | 0.199 | 0.69 |
| Mongolia | 1961 | 3.107 | 0.019 | 0.023 | 0.437 | 0.774 | 0.126 | 0.036 | 0.382 |
| | 2022 | 3.98 | 0.247 | 0.073 | 1.21 | 0.872 | 0.733 | 0.494 | 0.615 |
| Montenegro | 2006 | 3.1 | 0.981 | 0.197 | 1.047 | 1.037 | 1.509 | 1.018 | 0.863 |
| | 2022 | 3.805 | 0.789 | 0.306 | 0.986 | 1.039 | 1.287 | 0.782 | 0.81 |
| Morocco | 1961 | 0.393 | 0.294 | 0.113 | 0.52 | 1.184 | 1.187 | 0.209 | 0.422 |
| | 2022 | 0.937 | 1.052 | 0.386 | 1.199 | 1.69 | 1.426 | 0.633 | 0.826 |
| Mozambique | 1961 | 0.17 | 0.256 | 0.293 | 0.237 | 1.283 | 0.283 | 0.091 | 0.341 |
| | 2022 | 0.288 | 0.228 | 0.243 | 0.769 | 1.349 | 0.402 | 0.408 | 0.489 |
| Myanmar | 1961 | 0.28 | 0.288 | 0.197 | 0.45 | 0.909 | 0.2 | 0.127 | 0.375 |
| | 2022 | 0.712 | 0.486 | 0.779 | 0.565 | 1.583 | 0.752 | 0.66 | 0.7 |
| Namibia | 1961 | 0.953 | 0.225 | 0.41 | 0.733 | 1.077 | 1.196 | 0.136 | 0.576 |
| | 2022 | 0.925 | 0.354 | 0.28 | 0.921 | 1.176 | 1.405 | 0.185 | 0.611 |
| Nauru | 2019 | 2.653 | 2.435 | 0.412 | 1.071 | 0.567 | 2.11 | 0.215 | 0.699 |
| | 2022 | 2.514 | 2.381 | 0.388 | 1.133 | 0.589 | 2.142 | 0.303 | 0.713 |
| Nepal | 1961 | 0.45 | 0.062 | 0.207 | 0.263 | 1.246 | 0.048 | 0.045 | 0.338 |
| | 2022 | 0.66 | 0.474 | 0.439 | 1.193 | 1.604 | 0.46 | 0.993 | 0.761 |
| Netherlands | 1961 | 2.363 | 0.544 | 0.107 | 2.18 | 0.842 | 2 | 0.473 | 0.661 |
| | 2009 | 3.343 | 0.825 | 0.223 | 1.547 | 0.719 | 1.961 | 0.791 | 0.76 |
| Netherlands (Kingdom of the) | 2010 | 3.14 | 0.888 | 0.189 | 1.706 | 0.752 | 2.02 | 0.667 | 0.749 |
| | 2022 | 3.596 | 1.26 | 0.279 | 2.001 | 0.762 | 1.936 | 0.721 | 0.794 |
| Netherlands Antilles (former) | 1961 | 2.33 | 0.55 | 0.16 | 1.44 | 1.026 | 1.465 | 0.282 | 0.665 |
| | 2009 | 2.493 | 0.419 | 0.063 | 0.717 | 0.991 | 2.109 | 0.455 | 0.607 |
| New Caledonia | 1961 | 1.96 | 1.644 | 0.053 | 0.85 | 1.065 | 1.352 | 0.282 | 0.698 |

| Country | Year | | | | | | | | |
|---|---|---|---|---|---|---|---|---|---|
| | 2022 | 2.288 | 0.887 | 0.226 | 1.322 | 0.918 | 1.211 | 0.45 | 0.747 |
| New Zealand | 1961 | 2.8 | 0.631 | 0.167 | 1.33 | 0.736 | 2.083 | 0.509 | 0.674 |
| | 2022 | 2.174 | 0.849 | 0.372 | 1.565 | 0.896 | 2.084 | 0.661 | 0.796 |
| Nicaragua | 1961 | 0.773 | 0.275 | 0.587 | 0.263 | 0.85 | 1.478 | 0.073 | 0.47 |
| | 2022 | 1.316 | 0.109 | 0.518 | 0.818 | 1.215 | 1.598 | 0.222 | 0.611 |
| Niger | 1961 | 0.49 | 0.081 | 0.36 | 0.167 | 1.08 | 0.096 | 0.118 | 0.369 |
| | 2022 | 0.391 | 0.323 | 1.975 | 0.446 | 1.339 | 0.255 | 0.935 | 0.682 |
| Nigeria | 1961 | 0.153 | 0.281 | 0.517 | 1.037 | 1.108 | 0.07 | 0.355 | 0.551 |
| | 2022 | 0.232 | 0.735 | 0.721 | 1.024 | 1.402 | 0.359 | 0.451 | 0.69 |
| North Korea | 1961 | 0.403 | 0.069 | 0.707 | 0.09 | 1.204 | 0.057 | 0.5 | 0.461 |
| | 2018 | 0.376 | 0.486 | 0.418 | 0.495 | 1.162 | 0.182 | 0.828 | 0.6 |
| North Macedonia | 1992 | 1.07 | 0.569 | 0.273 | 0.817 | 0.995 | 0.922 | 1.027 | 0.776 |
| | 2022 | 1.63 | 0.982 | 0.435 | 1.64 | 0.973 | 1.554 | 1.86 | 0.898 |
| Norway | 1961 | 2.56 | 0.569 | 0.113 | 1.323 | 0.821 | 1.913 | 0.282 | 0.631 |
| | 2022 | 2.614 | 0.796 | 0.822 | 1.817 | 0.973 | 1.345 | 0.541 | 0.855 |
| Oman | 1990 | 1.29 | 1.538 | 0.123 | 0.9 | 0.917 | 1.004 | 0.673 | 0.769 |
| | 2022 | 1.984 | 2.924 | 0.391 | 1.681 | 0.69 | 1.24 | 1.48 | 0.847 |
| Pakistan | 1961 | 0.78 | 0.181 | 0.48 | 0.383 | 0.972 | 0.943 | 0.1 | 0.483 |
| | 2022 | 1.445 | 0.287 | 0.162 | 1.34 | 0.982 | 1.453 | 0.251 | 0.614 |
| Panama | 1961 | 0.903 | 0.675 | 0.22 | 0.533 | 0.993 | 1.183 | 0.155 | 0.58 |
| | 2022 | 1.756 | 0.837 | 0.268 | 1.67 | 1.16 | 1.11 | 0.306 | 0.735 |
| Papua New Guinea | 2010 | 1.082 | 3.401 | 0.039 | 0.704 | 0.747 | 0.386 | 0.517 | 0.668 |
| | 2022 | 0.936 | 2.942 | 0.043 | 0.768 | 0.903 | 0.461 | 0.439 | 0.682 |
| Paraguay | 1961 | 1.267 | 1.056 | 0.307 | 0.55 | 0.96 | 0.587 | 0.3 | 0.686 |
| | 2022 | 1.228 | 0.552 | 0.407 | 1.285 | 1.463 | 0.791 | 0.391 | 0.725 |
| Peru | 1961 | 0.643 | 0.338 | 0.287 | 0.62 | 1.044 | 1.2 | 0.309 | 0.533 |
| | 2022 | 1.629 | 1.246 | 0.32 | 0.821 | 1.212 | 0.87 | 0.509 | 0.775 |
| Philippines | 1961 | 0.62 | 1.069 | 0.063 | 0.263 | 0.975 | 0.613 | 0.455 | 0.563 |
| | 2022 | 0.965 | 0.687 | 0.134 | 0.769 | 1.558 | 1.001 | 0.45 | 0.668 |
| Poland | 1961 | 2.027 | 0.15 | 0.057 | 1.263 | 1.546 | 1.378 | 0.482 | 0.615 |
| | 2022 | 2.897 | 0.596 | 0.176 | 2.209 | 1.033 | 2.122 | 0.706 | 0.746 |

| Country | Year | | | | | | | | |
|---|---|---|---|---|---|---|---|---|---|
| Portugal | 1961 | 0.94 | 0.744 | 0.38 | 0.96 | 1.028 | 0.796 | 0.555 | 0.763 |
| | 2022 | 2.923 | 0.991 | 0.235 | 2.139 | 0.935 | 1.096 | 0.706 | 0.811 |
| Qatar | 2019 | 2.076 | 1.011 | 0.509 | 1.547 | 1.305 | 1.141 | 1.148 | 0.918 |
| | 2022 | 2.136 | 1.09 | 0.57 | 1.698 | 1.24 | 1.219 | 1.192 | 0.928 |
| Republic of Korea | 1961 | 0.16 | 0.038 | 0.24 | 0.057 | 1.595 | 0.078 | 0.536 | 0.338 |
| | 2022 | 1.906 | 0.444 | 0.461 | 2.244 | 0.999 | 2.066 | 1.719 | 0.817 |
| Republic of Moldova | 1992 | 1.773 | 0.456 | 0.153 | 0.51 | 1.106 | 1.026 | 0.564 | 0.614 |
| | 2022 | 1.575 | 1.054 | 0.422 | 1.348 | 0.911 | 1.766 | 0.742 | 0.846 |
| Romania | 1961 | 1.177 | 0.319 | 0.237 | 0.69 | 1.648 | 0.461 | 0.464 | 0.618 |
| | 2022 | 2.705 | 0.852 | 0.197 | 1.975 | 1.175 | 1.344 | 0.687 | 0.789 |
| Russian Federation | 1992 | 1.897 | 0.262 | 0.1 | 1.137 | 1.228 | 1.417 | 0.5 | 0.644 |
| | 2022 | 2.556 | 0.628 | 0.174 | 1.751 | 1.18 | 1.745 | 0.636 | 0.74 |
| Rwanda | 1961 | 0.103 | 3.394 | 1.093 | 0.037 | 0.697 | 0 | 0.155 | 0.499 |
| | 2022 | 0.248 | 1.388 | 0.989 | 0.413 | 1.197 | 0.59 | 0.284 | 0.656 |
| Saint Kitts and Nevis | 1961 | 0.627 | 0.631 | 0.093 | 0.83 | 0.608 | 1.665 | 0.073 | 0.477 |
| | 2022 | 2.044 | 1.198 | 0.344 | 1.636 | 0.7 | 1.977 | 0.249 | 0.715 |
| Saint Lucia | 1961 | 0.733 | 1.794 | 0.047 | 0.543 | 0.54 | 1.213 | 0.064 | 0.488 |
| | 2022 | 2.247 | 1.023 | 0.29 | 0.788 | 0.793 | 1.696 | 0.268 | 0.69 |
| Saint Vincent and the Grenadines | 1961 | 0.587 | 1.144 | 0.083 | 0.497 | 0.775 | 1.339 | 0.045 | 0.498 |
| | 2022 | 2.48 | 1.192 | 0.441 | 1.232 | 0.798 | 1.238 | 0.403 | 0.774 |
| Samoa | 1961 | 0.957 | 4.056 | 0 | 0.527 | 0.572 | 0.883 | 0.018 | 0.679 |
| | 2022 | 2.13 | 2.67 | 0.062 | 1.266 | 0.992 | 1.517 | 0.156 | 0.702 |
| Sao Tome and Principe | 1961 | 0.263 | 4.125 | 0.573 | 0.163 | 0.942 | 0.574 | 0.109 | 0.509 |
| | 2022 | 0.584 | 0.496 | 0.082 | 1.138 | 1.247 | 0.849 | 0.192 | 0.559 |
| Saudi Arabia | 1961 | 0.403 | 1.112 | 0.107 | 0.257 | 1.042 | 0.287 | 0.245 | 0.502 |

| Country | Year | | | | | | | | |
|---|---|---|---|---|---|---|---|---|---|
| | 2022 | 1.402 | 1.681 | 0.253 | 1.666 | 1.292 | 1.373 | 0.636 | 0.815 |
| Senegal | 1961 | 0.54 | 0.119 | 0.51 | 0.673 | 1.312 | 0.843 | 0.145 | 0.498 |
| | 2022 | 0.595 | 0.178 | 0.685 | 1.288 | 1.497 | 0.772 | 0.922 | 0.73 |
| Serbia | 2006 | 1.893 | 0.844 | 0.397 | 1.057 | 0.903 | 1.296 | 0.782 | 0.821 |
| | 2022 | 2.595 | 1.42 | 0.453 | 1.108 | 1.493 | 0.517 | 0.648 | 0.85 |
| Serbia and Montenegro | 1992 | 2.387 | 0.662 | 0.317 | 1.993 | 0.916 | 0.861 | 0.636 | 0.755 |
| | 2005 | 2.48 | 0.588 | 0.347 | 1.79 | 0.613 | 1.283 | 0.755 | 0.717 |
| Seychelles | 2010 | 1.338 | 0.743 | 0.222 | 0.57 | 1.298 | 1.595 | 0.577 | 0.685 |
| | 2022 | 2.105 | 1.639 | 0.361 | 0.763 | 1.093 | 1.152 | 0.915 | 0.84 |
| Sierra Leone | 1961 | 0.213 | 0.181 | 0.69 | 1.32 | 0.791 | 0.357 | 0.382 | 0.543 |
| | 2022 | 0.396 | 0.231 | 0.436 | 0.828 | 1.431 | 0.262 | 0.37 | 0.543 |
| Slovakia | 1993 | 1.927 | 0.338 | 0.193 | 1.6 | 0.897 | 1.383 | 0.455 | 0.647 |
| | 2022 | 1.851 | 0.511 | 0.173 | 3.108 | 0.729 | 1.138 | 0.49 | 0.65 |
| Slovenia | 1992 | 1.7 | 0.394 | 0.12 | 1.543 | 0.947 | 0.635 | 0.327 | 0.631 |
| | 2022 | 2.085 | 0.998 | 0.234 | 1.479 | 1.015 | 1.238 | 0.641 | 0.812 |
| Solomon Islands | 1961 | 0.537 | 1.725 | 0.257 | 0.14 | 1.347 | 0.274 | 0.136 | 0.512 |
| | 2022 | 0.573 | 1.146 | 0.273 | 0.661 | 1.243 | 0.698 | 0.072 | 0.596 |
| Somalia | 2010 | 1.268 | 0.157 | 0.144 | 0.372 | 0.621 | 1.006 | 0.051 | 0.391 |
| | 2022 | 0.879 | 0.161 | 0.162 | 0.47 | 0.665 | 2.434 | 0.086 | 0.404 |
| South Africa | 1961 | 1.197 | 0.181 | 0.143 | 0.627 | 1.335 | 1.665 | 0.336 | 0.548 |
| | 2022 | 1.501 | 0.279 | 0.119 | 1.553 | 1.09 | 1.27 | 0.295 | 0.616 |
| South Sudan | 2019 | 1.692 | 0.324 | 1.279 | 0.569 | 0.937 | 0.39 | 0.322 | 0.692 |
| | 2022 | 1.727 | 0.308 | 1.244 | 0.588 | 0.993 | 0.393 | 0.322 | 0.702 |
| Spain | 1961 | 1.087 | 0.456 | 0.447 | 1.253 | 1.118 | 0.87 | 0.955 | 0.81 |
| | 2022 | 2.881 | 0.679 | 0.424 | 2.519 | 0.797 | 1.438 | 0.744 | 0.774 |
| Sri Lanka | 1961 | 0.29 | 1.781 | 0.237 | 0.297 | 1.107 | 0.809 | 0.191 | 0.502 |
| | 2022 | 0.677 | 0.918 | 0.297 | 0.344 | 1.552 | 1.461 | 0.487 | 0.62 |
| Sudan | 2012 | 1.177 | 0.841 | 0.513 | 0.676 | 1.159 | 1.468 | 0.514 | 0.757 |
| | 2022 | 0.978 | 0.725 | 0.803 | 0.953 | 1.125 | 1.481 | 0.525 | 0.831 |
| Sudan (former) | 1961 | 0.927 | 0.362 | 0.38 | 0.457 | 0.724 | 0.522 | 0.2 | 0.508 |
| | 2009 | 1.703 | 0.575 | 0.44 | 0.593 | 0.953 | 1.057 | 0.455 | 0.669 |
| Suriname | 1961 | 0.773 | 0.188 | 0.187 | 0.733 | 0.844 | 1.283 | 0.118 | 0.474 |
| | 2022 | 1.226 | 0.488 | 0.16 | 1.044 | 1.005 | 2.12 | 0.331 | 0.663 |

| Country | Year | | | | | | | | |
|---|---|---|---|---|---|---|---|---|---|
| Sweden | 1961 | 2.483 | 0.531 | 0.08 | 2.11 | 0.657 | 1.961 | 0.209 | 0.58 |
| | 2022 | 2.845 | 0.589 | 0.277 | 2.177 | 0.865 | 1.473 | 0.665 | 0.733 |
| Switzerland | 1961 | 2.89 | 0.981 | 0.227 | 1.797 | 0.986 | 2.261 | 0.464 | 0.776 |
| | 2022 | 2.99 | 0.81 | 0.348 | 2.384 | 0.816 | 1.61 | 0.688 | 0.777 |
| Syrian Arab Republic | 2010 | 0.957 | 0.615 | 0.564 | 1.937 | 1.318 | 1.514 | 0.525 | 0.777 |
| | 2022 | 0.899 | 0.681 | 0.435 | 1.216 | 1.104 | 0.582 | 0.545 | 0.76 |
| Tajikistan | 1992 | 0.833 | 0.231 | 0.037 | 0.81 | 1.122 | 0.474 | 0.627 | 0.59 |
| | 2022 | 1.421 | 0.427 | 0.221 | 1.251 | 1.215 | 0.817 | 1.66 | 0.775 |
| Thailand | 1961 | 0.523 | 1.231 | 0.083 | 0.117 | 1.236 | 0.23 | 0.282 | 0.501 |
| | 2022 | 0.995 | 0.618 | 0.207 | 0.928 | 1.24 | 2.11 | 0.315 | 0.677 |
| Timor-Leste | 1961 | 1.173 | 0.506 | 0.23 | 0.137 | 1.008 | 0.03 | 0.173 | 0.508 |
| | 2022 | 0.775 | 0.167 | 0.333 | 0.944 | 1.188 | 0.707 | 0.232 | 0.575 |
| Togo | 1961 | 0.18 | 0.125 | 0.36 | 0.47 | 1.415 | 0.096 | 0.136 | 0.379 |
| | 2022 | 0.29 | 0.064 | 0.479 | 1.029 | 1.573 | 0.61 | 0.206 | 0.507 |
| Tonga | 2019 | 3.026 | 1.136 | 0.294 | 0.895 | 0.94 | 0.804 | 1.388 | 0.855 |
| | 2022 | 2.893 | 1.159 | 0.301 | 1.062 | 0.869 | 0.858 | 1.501 | 0.862 |
| Trinidad and Tobago | 1961 | 1.023 | 0.406 | 0.293 | 0.853 | 0.978 | 1.57 | 0.118 | 0.608 |
| | 2022 | 1.537 | 0.576 | 0.385 | 1.683 | 1.02 | 1.76 | 0.355 | 0.719 |
| Tunisia | 1961 | 0.513 | 0.444 | 0.127 | 0.773 | 1.224 | 1.017 | 0.436 | 0.549 |
| | 2022 | 1.295 | 1.234 | 0.499 | 1.809 | 1.386 | 1.495 | 1.483 | 0.916 |
| Turkey | 1961 | 1.327 | 1.312 | 0.463 | 0.897 | 1.486 | 0.296 | 0.873 | 0.872 |
| | 2022 | 2.257 | 1.317 | 0.777 | 2.09 | 1.299 | 1.27 | 1.444 | 0.963 |
| Turkmenistan | 1992 | 1.423 | 0.281 | 0.007 | 0.987 | 1.301 | 0.957 | 0.482 | 0.626 |
| | 2022 | 2.115 | 0.591 | 0.151 | 0.743 | 1.302 | 0.808 | 0.849 | 0.722 |
| Tuvalu | 2019 | 1.729 | 3.433 | 0.041 | 0.7 | 0.809 | 2.198 | 0.431 | 0.664 |
| | 2022 | 1.626 | 3 | 0.041 | 0.687 | 0.947 | 2.13 | 0.433 | 0.685 |
| Uganda | 1961 | 0.46 | 0.281 | 1.227 | 0.19 | 1.175 | 0.474 | 0.118 | 0.508 |
| | 2022 | 0.496 | 0.011 | 0.884 | 0.757 | 1.118 | 0.558 | 0.205 | 0.559 |
| Ukraine | 1992 | 2.02 | 0.319 | 0.163 | 1.31 | 1.397 | 2.165 | 0.509 | 0.665 |
| | 2022 | 1.936 | 0.442 | 0.118 | 1.107 | 1.096 | 1.438 | 1.067 | 0.76 |

| Country | Year | | | | | | | | |
|---|---|---|---|---|---|---|---|---|---|
| United Arab Emirates | 1961 | 2.343 | 0.419 | 0.013 | 1.463 | 1.142 | 1.504 | 0.036 | 0.578 |
| | 2022 | 1.937 | 0.879 | 0.782 | 2.327 | 1.091 | 1.02 | 0.756 | 0.903 |
| United Kingdom | 1961 | 2.96 | 0.488 | 0.15 | 1.933 | 0.814 | 2.191 | 0.355 | 0.634 |
| | 2022 | 2.727 | 0.784 | 0.222 | 2.027 | 0.967 | 1.229 | 0.692 | 0.777 |
| United Republic of Tanzania | 1961 | 0.407 | 0.231 | 0.37 | 0.227 | 1.065 | 0.296 | 0.3 | 0.422 |
| | 2022 | 0.577 | 0.626 | 0.868 | 0.818 | 1.19 | 0.496 | 0.234 | 0.687 |
| United States of America | 1961 | 2.7 | 0.531 | 0.283 | 1.59 | 0.62 | 2.243 | 0.582 | 0.669 |
| | 2022 | 3.263 | 0.964 | 0.456 | 2.891 | 0.793 | 2.625 | 0.662 | 0.813 |
| Uruguay | 1961 | 3.463 | 0.394 | 0.09 | 0.873 | 0.763 | 1.63 | 0.236 | 0.559 |
| | 2022 | 2.727 | 0.7 | 0.157 | 1.557 | 1.018 | 1.918 | 0.489 | 0.724 |
| USSR | 1961 | 1.73 | 0.169 | 0.15 | 0.823 | 1.528 | 1.348 | 0.382 | 0.587 |
| | 1991 | 2.197 | 0.306 | 0.117 | 1.113 | 1.23 | 1.53 | 0.473 | 0.649 |
| Uzbekistan | 1992 | 1.407 | 0.281 | 0.023 | 1.083 | 1.447 | 0.543 | 0.864 | 0.695 |
| | 2022 | 2.597 | 1.078 | 0.127 | 1.139 | 1.239 | 1.07 | 1.832 | 0.854 |
| Vanuatu | 1961 | 1.407 | 2.306 | 0.073 | 0.563 | 1.091 | 0.548 | 0.255 | 0.649 |
| | 2022 | 0.979 | 3.279 | 0.348 | 0.975 | 1.155 | 0.62 | 0.313 | 0.769 |
| Venezuela | 1961 | 1.003 | 1.181 | 0.297 | 0.823 | 0.768 | 1.526 | 0.109 | 0.666 |
| | 2022 | 1.106 | 0.594 | 0.351 | 1.354 | 0.856 | 1.32 | 0.314 | 0.686 |
| Viet Nam | 1961 | 0.417 | 0.306 | 0.113 | 0.09 | 1.363 | 0.191 | 0.273 | 0.366 |
| | 2022 | 1.472 | 0.766 | 0.609 | 0.539 | 1.435 | 1.168 | 1.213 | 0.819 |
| Yemen | 1961 | 0.36 | 0.325 | 0.223 | 0.273 | 1.165 | 0.457 | 0.091 | 0.379 |
| | 2022 | 0.411 | 0.393 | 0.225 | 0.561 | 1.135 | 1.077 | 0.181 | 0.462 |
| Yugoslav SFR | 1961 | 1.117 | 0.588 | 0.317 | 0.9 | 1.656 | 0.835 | 0.345 | 0.692 |
| | 1991 | 1.793 | 0.494 | 0.207 | 1.78 | 1.468 | 1.474 | 0.518 | 0.703 |
| Zambia | 1961 | 0.39 | 0.088 | 0.41 | 0.127 | 1.435 | 0.265 | 0.191 | 0.368 |
| | 2022 | 0.462 | 0.039 | 0.464 | 0.701 | 1.277 | 0.531 | 0.149 | 0.469 |
| Zimbabwe | 1961 | 0.637 | 0.112 | 0.407 | 0.183 | 1.344 | 0.426 | 0.109 | 0.408 |
| | 2022 | 1.238 | 0.158 | 0.134 | 1.184 | 0.868 | 0.767 | 0.088 | 0.541 |

Table S3. Regional HDBI and food supply as percentage of reference intake (1961 and 2022).

| Region | Year | ASF | Fruits | LNS | Oils & Fats | Starchy Staples | Sugars | Vegetables | HDBI |
|---|---|---|---|---|---|---|---|---|---|
| East Asia & Pacific | 1961 | 0.262 | 0.171 | 0.422 | 0.161 | 1.067 | 0.229 | 0.492 | 0.418 |
| | 2022 | 1.727 | 0.832 | 0.524 | 0.945 | 1.419 | 0.725 | 2.057 | 0.883 |
| Europe & Central Asia | 1961 | 1.877 | 0.438 | 0.173 | 1.297 | 1.238 | 1.374 | 0.482 | 0.682 |
| | 2022 | 2.720 | 0.770 | 0.272 | 2.122 | 0.973 | 1.584 | 0.794 | 0.802 |
| Latin America & Caribbean | 1961 | 1.063 | 0.547 | 0.488 | 0.547 | 0.957 | 1.538 | 0.222 | 0.627 |
| | 2022 | 2.147 | 0.915 | 0.357 | 1.668 | 1.009 | 1.685 | 0.402 | 0.779 |
| North America | 1961 | 2.677 | 0.537 | 0.270 | 1.597 | 0.622 | 2.226 | 0.573 | 0.667 |
| | 2022 | 3.168 | 0.954 | 0.478 | 2.904 | 0.799 | 2.550 | 0.671 | 0.817 |
| South Asia | 1961 | 0.340 | 0.306 | 0.647 | 0.380 | 1.151 | 0.787 | 0.209 | 0.480 |
| | 2022 | 0.911 | 0.585 | 0.534 | 1.026 | 1.226 | 0.942 | 0.621 | 0.775 |
| Sub-Saharan Africa | 1961 | 0.441 | 0.285 | 0.503 | 0.502 | 1.184 | 0.354 | 0.242 | 0.495 |
| | 2022 | 0.499 | 0.422 | 0.625 | 0.784 | 1.367 | 0.535 | 0.313 | 0.607 |
| Western Asia & North Africa | 1961 | 0.714 | 0.661 | 0.278 | 0.608 | 1.164 | 0.650 | 0.525 | 0.631 |
| | 2022 | 1.271 | 1.051 | 0.429 | 1.343 | 1.371 | 1.240 | 0.903 | 0.889 |
| World* | 1961 | 0.893 | 0.338 | 0.410 | 0.620 | 1.091 | 0.848 | 0.391 | 0.609 |
| | 2022 | 1.519 | 0.734 | 0.492 | 1.248 | 1.246 | 1.046 | 1.020 | 0.871 |
| Average of all countries** | 1961 | 1.031 | 0.632 | 0.296 | 0.697 | 1.068 | 0.974 | 0.299 | 0.548 |
| | 2022 | 1.691 | 0.856 | 0.390 | 1.331 | 1.103 | 1.249 | 0.634 | 0.717 |

* Denotes average values calculated from pooled global food supply.
** Denotes average values calculated from national data.

Table S4. IMPACT projections for regional HDBI and food supply as percentage of reference intake, increased investments and reference scenario (2010-2050).

| Region | Scenario | Year | ASF | Fruits | LNS | Oils & Fats | Starchy Staples | Sugars | Veg. | HDBI |
|---|---|---|---|---|---|---|---|---|---|---|
| East Asia & Pacific | Incr. Investments | 2010 | 1.241 | 0.968 | 0.343 | 1.062 | 1.145 | 1.211 | 0.52 | 0.67 |
| | | 2020 | 1.393 | 1.043 | 0.362 | 1.138 | 1.168 | 1.319 | 0.58 | 0.7 |
| | | 2030 | 1.538 | 1.132 | 0.385 | 1.201 | 1.225 | 1.401 | 0.61 | 0.73 |
| | | 2040 | 1.614 | 1.178 | 0.389 | 1.22 | 1.226 | 1.446 | 0.62 | 0.73 |
| | | 2050 | 1.665 | 1.211 | 0.391 | 1.231 | 1.211 | 1.478 | 0.63 | 0.74 |
| | Reference Scenario | 2010 | 1.241 | 0.968 | 0.343 | 1.062 | 1.145 | 1.211 | 0.52 | 0.67 |
| | | 2020 | 1.388 | 1.036 | 0.36 | 1.135 | 1.154 | 1.315 | 0.58 | 0.7 |
| | | 2030 | 1.499 | 1.092 | 0.37 | 1.184 | 1.155 | 1.391 | 0.61 | 0.72 |
| | | 2040 | 1.564 | 1.13 | 0.371 | 1.201 | 1.144 | 1.434 | 0.62 | 0.73 |
| | | 2050 | 1.613 | 1.162 | 0.373 | 1.211 | 1.128 | 1.465 | 0.63 | 0.73 |
| Europe & Central Asia | Incr. Investments | 2010 | 2.467 | 0.69 | 0.216 | 1.797 | 0.96 | 1.495 | 0.83 | 0.76 |
| | | 2020 | 2.492 | 0.721 | 0.221 | 1.801 | 0.968 | 1.548 | 0.87 | 0.77 |
| | | 2030 | 2.539 | 0.755 | 0.231 | 1.829 | 1.011 | 1.592 | 0.9 | 0.78 |
| | | 2040 | 2.564 | 0.779 | 0.234 | 1.847 | 1.019 | 1.625 | 0.9 | 0.78 |
| | | 2050 | 2.584 | 0.798 | 0.237 | 1.859 | 1.016 | 1.645 | 0.9 | 0.79 |
| | Reference Scenario | 2010 | 2.467 | 0.69 | 0.216 | 1.797 | 0.96 | 1.495 | 0.83 | 0.76 |
| | | 2020 | 2.488 | 0.72 | 0.22 | 1.799 | 0.959 | 1.546 | 0.87 | 0.77 |
| | | 2030 | 2.511 | 0.751 | 0.225 | 1.821 | 0.962 | 1.589 | 0.9 | 0.77 |
| | | 2040 | 2.53 | 0.775 | 0.227 | 1.839 | 0.961 | 1.623 | 0.91 | 0.78 |
| | | 2050 | 2.548 | 0.793 | 0.229 | 1.85 | 0.955 | 1.643 | 0.91 | 0.78 |
| Latin America & Caribbean | Incr. Investments | 2010 | 1.496 | 0.742 | 0.363 | 1.184 | 1.023 | 1.839 | 0.37 | 0.71 |
| | | 2020 | 1.554 | 0.788 | 0.379 | 1.221 | 1.031 | 1.926 | 0.41 | 0.72 |
| | | 2030 | 1.627 | 0.828 | 0.404 | 1.26 | 1.065 | 1.981 | 0.43 | 0.74 |
| | | 2040 | 1.662 | 0.852 | 0.418 | 1.284 | 1.064 | 2.02 | 0.44 | 0.75 |
| | | 2050 | 1.692 | 0.874 | 0.429 | 1.31 | 1.054 | 2.062 | 0.45 | 0.76 |
| | Reference Scenario | 2010 | 1.496 | 0.742 | 0.363 | 1.184 | 1.023 | 1.839 | 0.37 | 0.71 |
| | | 2020 | 1.547 | 0.786 | 0.377 | 1.219 | 1.022 | 1.921 | 0.41 | 0.72 |
| | | 2030 | 1.585 | 0.819 | 0.391 | 1.251 | 1.016 | 1.975 | 0.43 | 0.73 |
| | | 2040 | 1.61 | 0.841 | 0.4 | 1.273 | 1.006 | 2.013 | 0.44 | 0.74 |
| | | 2050 | 1.638 | 0.862 | 0.411 | 1.298 | 0.994 | 2.054 | 0.45 | 0.75 |
| North America | Incr. Investments | 2010 | 2.611 | 0.816 | 0.479 | 2.714 | 0.841 | 2.276 | 0.78 | 0.82 |
| | | 2020 | 2.616 | 0.828 | 0.485 | 2.612 | 0.841 | 2.313 | 0.82 | 0.83 |
| | | 2030 | 2.651 | 0.843 | 0.506 | 2.565 | 0.876 | 2.333 | 0.85 | 0.84 |
| | | 2040 | 2.675 | 0.858 | 0.511 | 2.581 | 0.886 | 2.361 | 0.87 | 0.85 |
| | | 2050 | 2.7 | 0.869 | 0.51 | 2.601 | 0.889 | 2.382 | 0.88 | 0.86 |
| | Reference Scenario | 2010 | 2.611 | 0.816 | 0.479 | 2.714 | 0.841 | 2.276 | 0.78 | 0.82 |
| | | 2020 | 2.611 | 0.827 | 0.482 | 2.612 | 0.832 | 2.31 | 0.82 | 0.83 |

| Region | Scenario | Year | | | | | | | | |
|---|---|---|---|---|---|---|---|---|---|---|
| | | 2030 | 2.617 | 0.838 | 0.484 | 2.568 | 0.828 | 2.337 | 0.86 | 0.83 |
| | | 2040 | 2.633 | 0.853 | 0.485 | 2.588 | 0.829 | 2.369 | 0.88 | 0.84 |
| | | 2050 | 2.656 | 0.863 | 0.484 | 2.608 | 0.829 | 2.39 | 0.89 | 0.84 |
| South Asia | Incr. Investments | 2010 | 0.601 | 0.426 | 0.273 | 0.754 | 1.369 | 0.771 | 0.38 | 0.56 |
| | | 2020 | 0.667 | 0.527 | 0.295 | 0.792 | 1.405 | 0.888 | 0.47 | 0.61 |
| | | 2030 | 0.749 | 0.628 | 0.325 | 0.848 | 1.511 | 0.977 | 0.57 | 0.65 |
| | | 2040 | 0.788 | 0.684 | 0.338 | 0.877 | 1.521 | 1.012 | 0.64 | 0.67 |
| | | 2050 | 0.816 | 0.726 | 0.348 | 0.905 | 1.502 | 1.029 | 0.71 | 0.69 |
| | Reference Scenario | 2010 | 0.601 | 0.426 | 0.273 | 0.754 | 1.369 | 0.771 | 0.38 | 0.56 |
| | | 2020 | 0.661 | 0.523 | 0.292 | 0.788 | 1.383 | 0.883 | 0.47 | 0.6 |
| | | 2030 | 0.711 | 0.607 | 0.308 | 0.826 | 1.394 | 0.96 | 0.56 | 0.64 |
| | | 2040 | 0.738 | 0.657 | 0.317 | 0.852 | 1.384 | 0.993 | 0.63 | 0.66 |
| | | 2050 | 0.761 | 0.694 | 0.326 | 0.878 | 1.361 | 1.008 | 0.69 | 0.67 |
| Sub-Saharan Africa | Incr. Investments | 2010 | 0.623 | 0.363 | 0.559 | 0.769 | 1.308 | 0.625 | 0.25 | 0.56 |
| | | 2020 | 0.686 | 0.397 | 0.584 | 0.799 | 1.344 | 0.646 | 0.28 | 0.58 |
| | | 2030 | 0.803 | 0.451 | 0.646 | 0.854 | 1.436 | 0.673 | 0.3 | 0.62 |
| | | 2040 | 0.887 | 0.485 | 0.668 | 0.881 | 1.43 | 0.695 | 0.32 | 0.64 |
| | | 2050 | 0.965 | 0.517 | 0.681 | 0.909 | 1.406 | 0.719 | 0.34 | 0.65 |
| | Reference Scenario | 2010 | 0.623 | 0.363 | 0.559 | 0.769 | 1.308 | 0.625 | 0.25 | 0.56 |
| | | 2020 | 0.679 | 0.393 | 0.577 | 0.794 | 1.322 | 0.643 | 0.28 | 0.58 |
| | | 2030 | 0.742 | 0.421 | 0.593 | 0.824 | 1.31 | 0.663 | 0.3 | 0.6 |
| | | 2040 | 0.798 | 0.443 | 0.601 | 0.842 | 1.282 | 0.679 | 0.31 | 0.61 |
| | | 2050 | 0.859 | 0.468 | 0.611 | 0.862 | 1.255 | 0.699 | 0.32 | 0.62 |
| Western Asia & North Africa | Incr. Investments | 2010 | 1.351 | 0.85 | 0.328 | 1.38 | 1.26 | 1.394 | 0.94 | 0.79 |
| | | 2020 | 1.405 | 0.868 | 0.343 | 1.416 | 1.275 | 1.515 | 0.99 | 0.8 |
| | | 2030 | 1.494 | 0.896 | 0.36 | 1.476 | 1.339 | 1.643 | 1.03 | 0.81 |
| | | 2040 | 1.538 | 0.906 | 0.37 | 1.504 | 1.343 | 1.731 | 1.05 | 0.82 |
| | | 2050 | 1.566 | 0.912 | 0.376 | 1.517 | 1.326 | 1.789 | 1.06 | 0.82 |
| | Reference Scenario | 2010 | 1.351 | 0.85 | 0.328 | 1.38 | 1.26 | 1.394 | 0.94 | 0.79 |
| | | 2020 | 1.398 | 0.867 | 0.342 | 1.412 | 1.261 | 1.509 | 0.99 | 0.8 |
| | | 2030 | 1.444 | 0.888 | 0.356 | 1.459 | 1.265 | 1.625 | 1.04 | 0.81 |
| | | 2040 | 1.475 | 0.897 | 0.365 | 1.484 | 1.256 | 1.707 | 1.06 | 0.81 |
| | | 2050 | 1.5 | 0.902 | 0.37 | 1.495 | 1.236 | 1.763 | 1.06 | 0.82 |
| Average of national HDBIs | Incr. Investments | 2010 | 1.425 | 0.653 | 0.372 | 1.237 | 1.144 | 1.236 | 0.55 | 0.68 |
| | | 2020 | 1.488 | 0.694 | 0.388 | 1.266 | 1.164 | 1.304 | 0.59 | 0.7 |
| | | 2030 | 1.578 | 0.744 | 0.419 | 1.312 | 1.227 | 1.362 | 0.62 | 0.72 |
| | | 2040 | 1.63 | 0.773 | 0.43 | 1.335 | 1.228 | 1.401 | 0.64 | 0.73 |
| | | 2050 | 1.673 | 0.798 | 0.438 | 1.355 | 1.214 | 1.432 | 0.65 | 0.74 |
| | Reference Scenario | 2010 | 1.425 | 0.653 | 0.372 | 1.237 | 1.144 | 1.236 | 0.55 | 0.68 |
| | | 2020 | 1.482 | 0.691 | 0.385 | 1.263 | 1.15 | 1.3 | 0.59 | 0.7 |
| | | 2030 | 1.534 | 0.726 | 0.396 | 1.295 | 1.147 | 1.353 | 0.62 | 0.71 |

| Year | | | | | | | | |
|------|-------|-------|-------|-------|-------|-------|------|------|
| 2040 | 1.572 | 0.75  | 0.402 | 1.315 | 1.134 | 1.39  | 0.64 | 0.72 |
| 2050 | 1.608 | 0.772 | 0.409 | 1.331 | 1.118 | 1.419 | 0.65 | 0.72 |

**Table S5. List of countries and country groups included in IMPACT modeling.**

| IMPACT country |
|---|
| * country or territory is included only as part of an aggregate group |
| † country group is excluded from Figure 4 |

- Afghanistan
- Angola
- Albania
- Argentina
- Armenia
- Australia
- Austria
- Azerbaijan
- Burundi
- Benin
- Burkina Faso
- Bangladesh
- Bulgaria
- Belarus
- Baltic States:
  - Estonia*
  - Lithuania*
  - Latvia*
- Belgium-Luxembourg
- Belize
- Bolivia
- Brazil
- Bhutan
- Botswana
- Central African Republic
- Canada
- Chile
- China Plus:
  - China*
  - Hong Kong*
  - Macao*
  - Taiwan*

Switzerland Plus:
- Switzerland*
- Liechtenstein*

Ivory Coast

Cameroon

Democratic Republic of Congo

Congo

Colombia

Other Caribbean†:
- Aruba*
- Anguilla*
- Netherlands Antilles (obsolete) *
- Antigua*
- Bonaire, Sint Eustatius, and Saba*
- Bahamas*
- St. Barthelemy*
- Barbados*
- Curacao*
- Cayman Islands*
- Dominica*
- Guadeloupe*
- Grenada*
- St. Kitts and Nevis*
- St. Lucia*
- Saint Martin*
- Montserrat*
- Martinique*
- Puerto Rico*
- Sint Maarten*
- Turks and Caicos Islands*
- Trinidad and Tobago*
- St. Vincent and Grenadines*
- British Virgin Islands*
- US Virgin Islands*

Costa Rica

Cuba

Cyprus
Czech Republic
Germany
Djibouti
Denmark
Dominican Republic
Algeria
Ecuador
Egypt
Eritrea
Ethiopia
Fiji
Finland Plus:
- Aland Islands*
- Finland*

France Plus:
- France*
- Monaco*

Gabon
Georgia
Ghana
Guinea
Gambia
Guinea-Bissau
Equatorial Guinea
Greece
Greenland
Guyanas South America†:
- French Guiana*
- Guyana*
- Suriname*

Guatemala
Honduras
Croatia
Haiti
Hungary

Indonesia
India
Ireland
Iran
Iraq
Iceland
Israel
Italy Plus:
    Italy*
    Malta*
    San Marino*
    Vatican City*
Jamaica
Jordan
Japan
Kazakhstan
Kenya
Kyrgyzstan
Cambodia
South Korea
Laos
Lebanon
Liberia
Libya
Sri Lanka
Lesotho
Moldova
Madagascar
Mexico
Mali
Myanmar
Mongolia
Morocco Plus:
    Morocco*
    Western Sahara*
Mozambique

Mauritania

Malawi

Malaysia

Namibia

Niger

Nigeria

Nicaragua

Netherlands

Norway

Nepal

New Zealand

Other Atlantic Ocean†:

- Bermuda*
- Bouvet Island*
- Cape Verde*
- Falkland Islands*
- Faroe Islands*
- South Georgia and South Sandwich Islands*
- Saint Helena, Ascension, and Tristan de Cunha*
- Svalbard and Jan Mayen*
- Saint Pierre and Miquelon*
- Sao Tome and Principe*

Other Balkans:

- Bosnia-Herzegovina*
- Macedonia (FYR) *
- Montenegro*
- Serbia*

Other Indian Ocean†:

- Southern Territories*
- Keeling Islands*
- Comoros*
- Christmas Island*
- Heard and McDonald Islands*
- British Indian Ocean Territory*
- Maldives*

    Mauritius*
    Mayotte*
    Reunion*
    Seychelles*
Other Pacific Ocean†:
    American Samoa*
    Cook Islands*
    Micronesia*
    Guam*
    Kiribati*
    Marshall Islands*
    Northern Mariana Islands*
    New Caledonia*
    Norfolk Island*
    Niue*
    Nauru*
    Pitcairn*
    Palau*
    French Polynesia*
    Tokelau*
    Tonga*
    Tuvalu*
    Minor Outlying Islands*
    Wallis and Futuna*
    Samoa*
Other Southeast Asia:
    Brunei*
    Singapore*
Pakistan
Panama
Peru
Philippines
Papua New Guinea
Poland
North Korea
Portugal

Paraguay
Occupied Palestinian Territory
Rest of Arab Peninsula:
    United Arab Emirates*
    Bahrain*
    Kuwait*
    Oman*
    Qatar*
Romania
Russia
Rwanda
Saudi Arabia
Sudan Plus:
    Sudan*
    South Sudan*
Senegal
Solomon Islands
Sierra Leone
El Salvador
Somalia
Spain Plus:
    Andorra*
    Spain*
    Gibraltar*
Slovakia
Slovenia
Sweden
Swaziland
Syria
Chad
Togo
Thailand
Tajikistan
Turkmenistan
Timor-L'este
Tunisia

Turkey
Tanzania
Uganda
Great Britain Plus:
    Great Britain*
    Guernsey*
    Isle of Man*
    Jersey*
Ukraine
Uruguay
United States
Uzbekistan
Venezuela
Vietnam
Vanuatu
Yemen
South Africa
Zambia
Zimbabwe